\documentstyle[epsfig,11pt]{article}

\begin{document}

\title{Quantum information processing in semiconductor nanostructures}

\author{John Henry Reina,$^{1,\ast}$ Luis Quiroga,$^{2,\dag}$ and Neil F. Johnson$^{1,\ddag}$\\
\footnotesize$^1$Physics Department, Clarendon Laboratory, Oxford
University, Oxford, OX1 3PU, United Kingdom\\
\footnotesize$^2$Departamento de F\'{\i}sica, Universidad de los Andes, Santaf\'{e}
de Bogot\'{a}, A.A. 4976, Colombia
}

\maketitle

\begin{abstract}
\footnote{Invited chapter for the Proceedings of the
ISI-Accademia dei Lincei Conference ``Conventional and non Conventional Computing
(Quantum and DNA)", to be published by Springer Verlag.}A major question for condensed matter physics is whether
a solid-state quantum computer can ever be built. Here we discuss two different schemes for quantum  information
processing using semiconductor nanostructures. First,  we show how optically driven coupled quantum dots can be
used to prepare maximally entangled Bell and Greenberger-Horne-Zeilinger states by varying the strength and
duration of selective light pulses. The setup allows us to perform an all-optical generation of the quantum
teleportation of an excitonic state in an  array of coupled quantum dots.
 Second, we
give a proposal for reliable implementation of quantum logic gates and long decoherence times 
in a quantum dots system based on nuclear
magnetic resonance (NMR), where the nuclear resonance is controlled by the ground state transitions 
of few-electron QDs in an external
magnetic field. The dynamical evolution of these systems in the presence of 
environmentally-induced 
decoherence effects is also discussed.
\end{abstract}

\vspace{0.2cm}

\begin{center}
\begin{tabular}{l c c}\small
& \small\ Phone \  & \small\ Fax \cr
$^{*}$\small\texttt{j.reina-estupinan@physics.ox.ac.uk} & \ (44 1865) 272257 \  & \ 272400\cr
$^\dag$\small\texttt{luis@anacaona.uniandes.edu.co} & \ (57 1) 2839514 \  & \ 2839514\cr
$^\ddag$\small\texttt{n.johnson@physics.ox.ac.uk} & \ (44 1865) 272287 \  & \  272400\cr

\end{tabular}
\end{center}

%\vspace{0.5cm}

\tableofcontents

\section{Introduction}

It has become increasingly clear that quantum mechanical 
principles are not just 
exotic theoretical statements but fundamental for a new technology 
of practical information 
processing \cite{PW}. Quantum computation, quantum 
cryptography and quantum teleportation represent exciting new arenas 
which 
exploit intrinsic quantum mechanical correlations. 

The discovery of algorithms for which a computer based on the 
principles of 
quantum mechanics \cite{Shor} should beat any traditional computer, 
has triggered intense 
research into realistic controllable quantum systems. 
Among the main areas 
involved in this active research field are ion traps \cite{Trapped Ions}, 
quantum electrodynamics 
cavities \cite{Cavity QED},  
nuclear magnetic 
resonance (NMR) \cite{NMR}, Josephson junctions \cite{Superconductors} and semiconductor quantum dots (QDs)
\cite{Semiconductors}.  The 
main challenge now 
is to identify a physical system with an appropriate internal dynamics and 
corresponding external driving forces which 
enables one to 
selectively manipulate quantum superpositions and entanglements. 
A fundamental 
requirement for the experimental realization of such proposals is the 
successful 
generation of highly entangled quantum states. In particular, coherent 
evolution 
of two quantum bits (qubits) in an entangled state of the Bell type is 
fundamental 
to both quantum cryptography and quantum teleportation. Maximally 
entangled 
states of three qubits, such as the so-called 
Greenberger-Horne-Zeilinger 
(GHZ) states \cite{GHZ}, are not only of intrinsic interest but are also 
of 
great practical importance in such proposals. 
Besides the capability to control and manipulate entanglement a 
great level of isolation from the environment is required to 
reach a full unitary evolution. 
Quantum information processing will be a reality when optimal control 
of 
quantum coherence 
in noisy environments  
can be achieved. 
The various 
communities typically rely on different hardware methodologies. 
Therefore, it is 
extremely important to clarify the underlying physics and limits for 
each type of 
physical realization of quantum information processing systems. 

In this chapter we discuss two possible strategies using 
semiconductor QDs \cite{Neil}. 
First, we review our main results on the optical generation and control 
of exciton\footnote{Excitons are electronic 
excitations 
which play a fundamental role 
in the optical properties of dielectric solids. They correspond to a 
bound 
state 
of one electron and one hole which can be created by light or can appear 
as 
a 
result of relaxation processes of free electrons and holes.}  
entangled states in coupled QDs by using 
a state-of-the-art semiconductor setup that enables us to generate 
reliable 
maximally entangled states of $N$ qubits, starting from suitably 
initialized 
states. 
As an application of these exciton 
maximally entangled states, 
a true solid-state teleportation protocol is proposed. 
We show that the role of phonons, at low temperatures, in the driven 
QD system does not necessarily amount to the loss 
of control over the system due to destruction of coherence. 
Second, we address the implementation of a solid state NMR-based quantum switch. 
We discuss how the so-called ``magic-number" transitions in few-electron 
QDs containing 
a nuclear spin impurity inside can be 
used to implement single qubit rotations and controlled$-$NOT (C$-$NOT) quantum  
gates. 
The basic setup 
consists of a nuclear 
spin-$\frac{1}{2}$ impurity 
placed at the center of a 2 electrons QD in the presence of an external 
perpendicular magnetic 
field $B$. In such a system, 
the nuclear magnetic resonance is controlled by the ground state 
transitions 
that arise as the 
$B$-field is changed: we show 
that the hyperfine coupling between the electrons and the nucleus can be 
changed and hence 
provide a mechanism for 
tuning the nuclear resonance frequency. 
Decoherence effects 
in systems of spin $\frac{1}{2}$ nuclei are expected to be minimal as nuclear  
spins are 
weakly 
coupled to their environment. Therefore, such spin systems 
are natural qubits for quantum 
information 
processing since 
they offer long decoherence times.  
Indeed they have been used in 
bulk liquid NMR experiments to perform some basic quantum algorithms 
like those of Deutsch \cite{mosca1} and Grover \cite{mosca2}. 
They have 
already been employed in some solid-state proposals, for example that of 
Ref. \cite{kane}Ê where a set of donor atoms (like P) is embedded in 
pure 
silicon. Here, the qubit is represented by the nuclear spin of the donor 
atom and single qubit and C$-$NOT operations might 
then 
be 
achieved between neighbor nuclei by attaching electric gates on top and 
between the donor atoms \cite{kane}. Another proposal \cite{privman} 
suggests controlling the 
hyperfine electron-nuclear interaction via the excitation of the 
electron 
gas in quantum Hall systems. Both of these 
proposals, however, require the attachment of electrodes or gates to the 
sample in order to manipulate the nuclear spin qubit. Such electrodes 
are likely to have an invasive effect on the coherent evolution of 
the qubit, thereby destroying quantum information. 
ÊIn the second part of 
this chapter, we propose 
a NMR solid-state based mechanism for quantum computation free from these 
shortcomings.  
The outline of this chapter is as follows: In Section 2 we give a 
detailed 
prescription for producing 
Êmaximally entangled exciton 
states of 
two 
and three semiconductor QDs. Section 3 considers the 
Êeffects of 
decoherence on the optical generation of such entangled states. In 
Section 
4, a protocol for 
teleporting the 
excitonic state of a quantum dot is proposed. In Section 5 we give a 
novel 
model for quantum logic 
with an NMR$-$based 
nanostructure switch. Concluding remarks are given in Section 6.

\section{Generation of maximally entangled exciton states in
optically driven quantum dots}

When two quantum dots are 
sufficiently close, there is a resonant energy transfer process originating from
the Coulomb interaction whereby an exciton can hop between dots \cite{Mahler}.
Experimental evidence of such energy transfers between quantum dots was reported
recently \cite{Bonadeo2}; the resonant process also plays a fundamental role
in biological and organic systems, and is commonly called the F\"{o}rster
process \cite{PT}. Unlike usual single-particle transport measurements, the
F\"{o}rster process does not require the physical transfer of the electron and
the hole, just their energy. Hence it is relatively insensitive to the effects of
impurities which lie between the dots. 

Here we show how the resonant transfer (F\"orster) interaction between spatially separated 
excitons can be exploited to produce maximally entangled states of two
(Bell) and three (GHZ)  optically driven QDs, starting from suitably initialized states. Previous
experiments have studied entangled states of trapped ions \cite{Trapped Ions}, photons \cite{photons}, 
and particle
spins in bulk liquid NMR \cite{Chuang}, but to our knowledge,  there is not such an scheme for 
producing deterministic
entanglement in a semiconductor nanostructures setup. In the proposal given here we exploit 
recent experimental 
results involving {\it coherent wavefunction control} of excitons in semiconductor quantum 
dots on the nanometer and
femtosecond scales \cite{Bonadeo2,Bonadeo1,axt,ChavezPirson}, i.e, the system requirements 
can be realized with
current experiments employing both ultrafast and near-field optical spectroscopy of 
quantum dots. 

%\subsection{Model and Results}

We denote by 0 (1) a zero-exciton (single exciton)
QD. We consider a system of
$N$ identical and equispaced QDs, containing no net charge,  which are radiated by 
long-wavelength
classical light (see Figure 1) in order to produce reliable generation of the 
maximally entangled states $\left| \Psi _{Bell}(\varphi)\right\rangle =
{\textstyle{1 \over \sqrt{2}}}(\left|
00\right\rangle  +e^{i\varphi }\left| 11\right\rangle )\;$and $\left|\Psi _{GHZ}(\varphi)\right\rangle =
{\textstyle{1 \over \sqrt{2}}}(\left| 000\right\rangle +e^{i\varphi }\left| 111\right\rangle )
$,$\;$for several different values of the phase factor $\varphi$. 
The formation of single excitons within the
individual QDs and their inter-dot hopping can be described in the 
frame of the {\it rotating wave approximation} (RWA) by
the Hamiltonian ($\hbar=1$) \cite{reina1}: 

\begin{equation}
H_{_\Lambda}=\Delta _{\omega }J_{z}+A(J_{+}+J_{-})+W(J^{2}-J_{z}^{2})\ \ .
\end{equation}
Here $\Delta _{\omega }\equiv\epsilon-\omega$ is the
detuning parameter, $\epsilon$ is the QD band gap, $W$ represents the 
interdot Coulomb interaction (F\"{o}rster
process), the subscript $\Lambda$ refers to the rotating frame (see below), and the operators
$J_+=\sum_{n=1}^{N} {e_{n}^{\dag}h_{n}^{\dag}}$,
$J_-=\sum_{n=1}^{N}{h_{n}e_{n}}$,
$J_z=\frac {1}{2} \sum_{n=1}^{N} {(e_{n}^{\dag}e_{n}-
h_{n}h_{n}^{\dag})}$,
with $e_{n}^{\dag}$ ($h_n^{\dag}$) describing the electron (hole)
creation operator
in the $n$'th QD. The \ $J_{i}-$operators obey the usual angular
momentum commutation relations $\left[J_{z},J_{\pm } \right] =\pm J_{\pm },\;\left[ J_{+},J_{-}\right] =2
J_{z},\;
$and$
\; [J^{2},J_{+}]=[J^{2},J_{-}]=[J^{2},J_{z}]=0,$ where $J^{2}\equiv
{\textstyle{1 \over 2}}
[J_{+}J_{-}+J_{-}J_{+}]+J_{z}^{^{2}}$.$\;$ We consider the situation of a 
laser pulse with central frequency
$\omega$ given by
$\xi (t)=Ae^{-i\omega t}$, where
$A$ gives the electron-photon coupling and the incident electric field strength. From
a practical point of view, parameters $A\;$and $\Delta _{\omega }$\ are adjustable in
the experiment to give control over the system of QDs. 
\begin{figure}
%\psdraft
\centerline{\epsfig{file=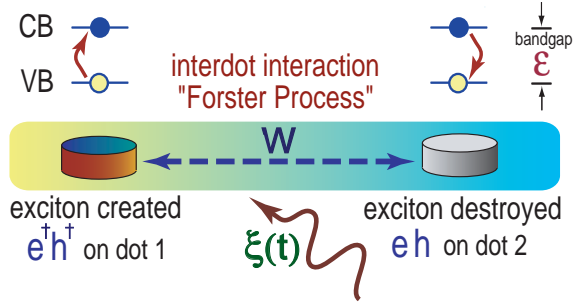,width=7.5cm}}
%\vspace{0.2cm}
\caption{\small Schematic of the optical setup for the $N=2$ QDs system. The identical QDs,
containing no net charge, are radiated with long-wavelength classical light
of central frequency
$\omega$, $\xi (t)=Ae^{-i\omega t}$. Formation of single excitons within the
individual QDs and their inter-dot hopping in the presence of the F\"{o}rster 
interaction are illustrated schematically.  The bandgap
$\epsilon$ as well as the conduction band (CB) and the valence band (VB) of 
the system are also shown.}
\label{figure1}
\end{figure}
Next, we discuss the main results obtained from the computation
of both analytical and numerical solutions for the time evolution of $H_{_\Lambda}$ \cite{reina1}. 
The solution to the 
quantum dynamical 
equation of motion of the system is equivalently given in terms of both the wave 
function and the density matrix
formalisms,  enabling us the strength and the length of the laser pulses required
for reliably generation of maximally entangled exciton states of two and three QDs.

\subsection{Unitary evolution and the wave function}

The total wave function of the excitonic system considered here, starting with
the initial condition $\left| \Psi (t=0)\right\rangle =\left| \Psi
_{0}\right\rangle $ (for any $N$), can be expressed as $\left| \Psi
(t)\right\rangle _{\Lambda }=
\mathop{\textstyle\sum}
_{k}C_{k}e^{-iE_{k}t}\left| \psi _{k}\right\rangle$, where 
$H\left| \psi _{k}\right\rangle =E_{k}\left| \psi
_{k}\right\rangle$ ($H$ is the Hamiltonian in the laboratory frame), and $\left| \psi _{k}\right\rangle =
\mathop{\textstyle\sum}
_{j}A_{kj}\left| M_{j}\right\rangle.$ As mentioned before, the subscript $\Lambda$ 
refers to the unitary
transformation which leads us from the laboratory frame to the rotating frame by using the rule
$\left| \Psi (t)\right\rangle _{\Lambda }=\Lambda ^{\dagger }(t)\left| \Psi
(t)\right\rangle_{S}$, with
$\Lambda=e^{-i\omega J_{Z}t}$ (subscript $S$ denotes Schr\"odinger picture). The normalization
coefficients $C_{k}$ depend on the chosen initial condition $%
\left| \Psi _{0}\right\rangle $:  by writing  $\left| \Psi (0)\right\rangle =
\mathop{\textstyle\sum}
_{k}\beta _{k}\left| M_{k}\right\rangle $ $\left( \beta _{k}=\left\langle
M_{k}\right. \left| \Psi (0)\right\rangle \right)$\ we see, from the above expansion
given for $\left| \Psi (t)\right\rangle _{\Lambda },$ that $\left|
\Psi (0)\right\rangle =
\mathop{\textstyle\sum}
_{k}C_{k}\left| \psi _{k}\right\rangle.$ Hence, the general expression for
the coefficients $C_{k}$ becomes $C_{k}=\left\langle \psi _{k}\right. \left|
\Psi (0)\right\rangle =
\mathop{\textstyle\sum}
_{j}\beta _{j}\left\langle \psi _{k}\right. \left| M_{j}\right\rangle =
\mathop{\textstyle\sum}
_{j}\beta _{j}A_{kj}^{\ast }.$
The matrix elements $A_{kj}$ must be
determined for each particular value of $N$, and $\left| M_{j}\right\rangle
\equiv \left| J,M_{j};q\right\rangle $, where $J$ can take the values
 $\frac{N}{2},\frac{N}{2}-1,...,\frac{1}{2}\;$or$
\;0$,$\;$and\ for\ each $J-$fixed value, we have the $2J+1$ different values 
$\;M=-\frac{N}{2},-\frac{N}{2}+1,...,\frac{N}{2}-1,
\frac{N}{2}.$  The label $q\;$is introduced to further distinguish the states:
$\ q=1,2,...,D_{J}$, where the multiplicity $D_{J}$, i.e. the number of states
having angular momentum $J$ and $M=J$, is given by $D_{J}=\frac{2J+1}{
J+\frac{N}{2}+1}{N \choose \frac{N}{2}+J}$. Hence, the total wave function in 
the rotating frame can be written as
\begin{equation}
\left| \Psi (t)\right\rangle _{\Lambda}=
\sum_{k}{\sum_{j}C_{k}A_{kj}e^{-iE_{k}t}\left| M_{j}\right\rangle} \ \ .  
\end{equation}
The eigenfunction given in Eq. (2) describes
any number of QDs. We only need to diagonalize a square matrix of side $2J+1$ for
each $J$. Every eigenvalue so obtained occurs $D_{J}$ times in the entire
spectrum. Next, we show how to generate highly excitonic entangled
states by solving the quantum equation of motion
associated with Eq. (2) for the cases $N=2$ and 3.

\subsubsection{Two coupled QDs and Bell states}

Here we give the light excitation procedure to obtain the
maximally entangled Bell states 
$|\Psi_{Bell}(\varphi)\rangle=|00\rangle+e^{i\varphi}|11\rangle$. The phase 
$\varphi$ determines the type of entangled state generated 
in the optical process. We
choose the basis of eigenstates of $J^2$ and $J_z$, 
 $\{\left| M_{1}\right\rangle \equiv \left|
J=1,M=-1\right\rangle \equiv \left| 0\right\rangle $, $\left|
M_{2}\right\rangle \equiv \left| J=1,M=0\right\rangle \equiv \left|
1\right\rangle$, $\left| M_{3}\right\rangle \equiv \left|
J=1,M=1\right\rangle \equiv \left| 2\right\rangle \}$, as an appropriate
representation for this problem. Here  $|0\rangle$ represents the vacuum for excitons,
$|1\rangle$ denotes the single-exciton state while $|2\rangle$ represents the
biexciton state. In the absence of light, we have that 
$E(J,M)=\Delta _{\omega }M+W[J(J+1)-M^{2}]$, so the energy levels of the system are
$E_{0}\equiv E(1,-1)=W-\Delta _{\omega }$, $E_{1}\equiv E(1,0)=2W$, and $E_{2}\equiv
E(1,1)=W+\Delta _{\omega }$. Next, consider the action of the
radiation pulse of light $\xi (t)$ over this pair of qubits at resonance, i.e.
$\Delta_{\omega}=0$. In 
this case, the new eigen-energies of the coupled system are: $
E_{0}=W,\;$and$\;E_{1,2}=
{\textstyle{1 \over 2}}
\Big(3W\pm \sqrt{16\left| A\right| ^{2}+W^{2}}\Big).$ Here we have assumed
that the decoherence processes are negligibly small over the time scale of the evolution 
(see next section). It is a straightforward exercise to compute the
explicit coefficients of Eq. (2) for both of the $J-$subspaces that span the Hilbert space
$SU(2)\otimes SU(2)$ ($N=2$) \cite{reina1}. Hence, the density
of probability $\wp (Bell)\;$for finding the entangled Bell state between vacuum and
biexciton states as a function of time for the initial
condition $\left| \Psi _{0}\right\rangle =\left| 0\right\rangle $ can be calculated as
\begin{equation}
\wp (Bell)=
{\textstyle{1 \over 2}}
\left| 
\mathop{\displaystyle\sum}
_{k}C_{k}\left(
A_{k1}+e^{i\varphi }A_{k3}\right)e^{-iE_{k}t}\right| ^{2}.  
\end{equation}
Results of the computation of Eq. (3) are shown in Fig. 2. 
Here we show several different
selective pulses of light  $\tau _{_{B}}$ that produce the entangled state  
$|\Psi_{Bell}(\varphi=0)\rangle$. 
In these figures, energies are given in terms of the
band gap $\epsilon $: $W=0.1$,
and (a)$\; A={\textstyle {1 \over 25}}$, (b)$\; A={\textstyle {1 \over
50}}$, (c)$\;A=10^{-2}$, and (d)$\;A=10^{-3}$. Here the energy $W$ is kept fixed while
the amplitude of the radiation pulse $A$ is varied. 
\begin{figure}
%\psdraft
\centerline{\epsfig{file=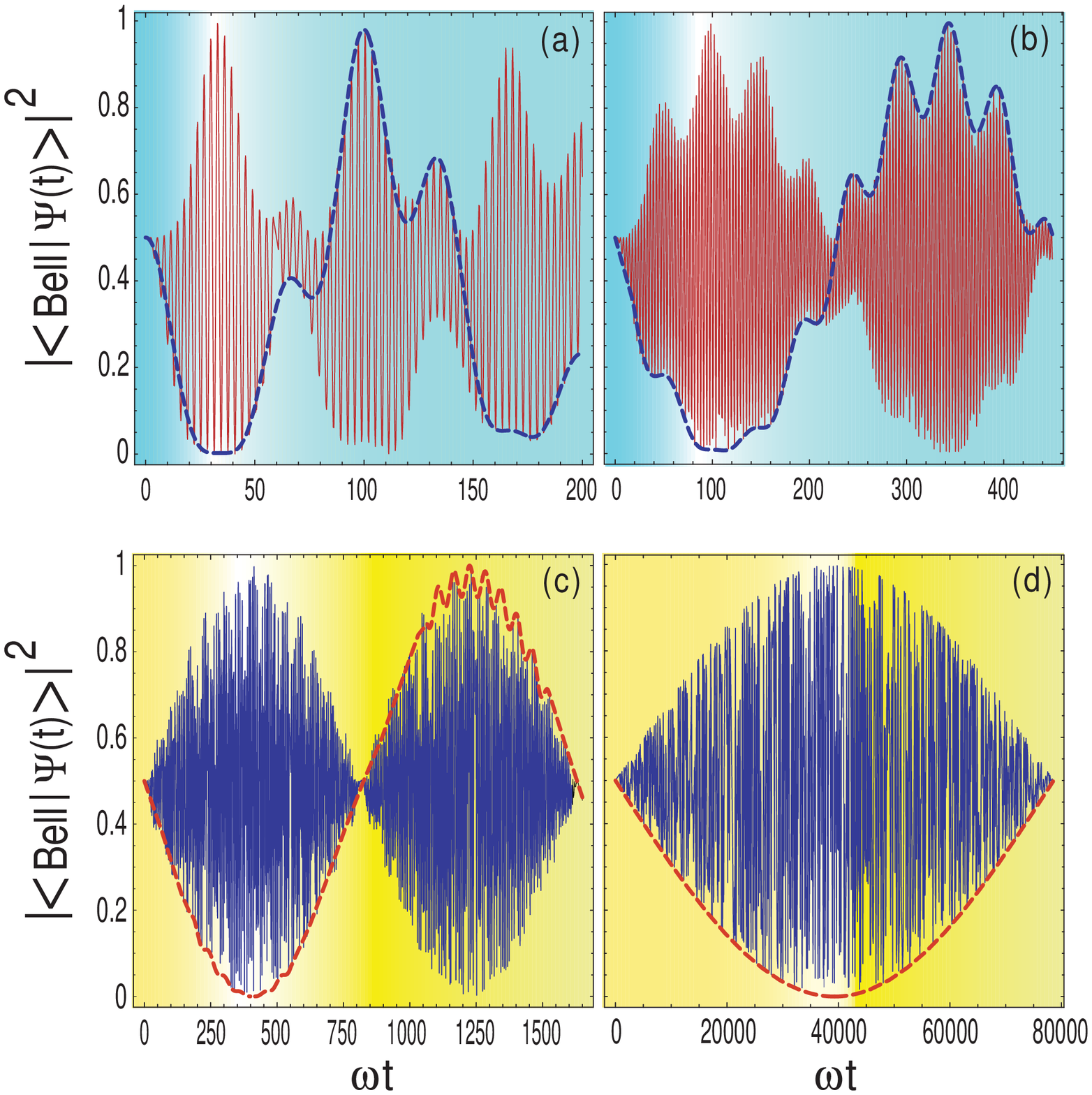,width=8.7cm,height=8.7cm,height=8.7cm}}
%\vspace{0.2cm}
\caption{\small{Generation of the Bell State $%\left| \Psi _{Bell}\right\rangle = 
{\textstyle{1 \over \sqrt{2}}}
(\left| 00\right\rangle +\left| 11\right\rangle )$.\ These pulses correspond
to the realization of the Hadamard gate followed by a quantum CNOT
gate. $W=0.1$,$\;\varphi =0,\;$
and (a)$\; A={\textstyle {1 \over 25}}$, (b)$\; A={\textstyle {1 \over
50}}$, (c)$\;A=10^{-2}$, and (d)$\;A=10^{-3}$. $\left| \Psi
(t)\right\rangle$ denotes the total wavefunction of the system at time $t$ in both laboratory
(solid curves) and rotating frames (dashed curves). The energy is in units of the
band gap $\epsilon $, and $\left| \Psi _{0}\right\rangle =\left|
0\right\rangle $.}}
\label{figure2}
\end{figure}
As a result of this, the
time $\tau _{_{B}}$ increases with diminishing incident 
field strength $A$ \cite{reina1}. We also consider another method for manipulating the length $\tau
_{_{B}}$: keeping $A$ fixed while varying $W$. In this case, the analysis
shows that for a fixed value of
$A$ the length $\tau _{_{B}}$ decreases with decreasing interaction strength 
$W$ \cite{reina1}. The latter procedure could be experimentally more expensive than the 
former since the variation 
of $W$ has to be tailored by changing the interdot distance and/or the radius of the dots. However, 
this method offers an interesting experimental possibility for studying the F\"orster mechanism.

Regarding the experimental generation of these Bell states, we suggest a 
consideration of wide-gap semiconductor QDs, like ZnSe based QDs, for instance.
For these materials, the band gap $\epsilon =2.8\;$ eV, which implies a resonant 
optical frequency $\omega
=4.3\times 10^{15}$ s$^{-1}$.  Femtosecond spectroscopy is currently available
for these  systems\cite{axt}. For a $\varphi=0$ or $2\pi
$ pulse, $W=0.1$ $\epsilon\;$and$\;A=0.04$ $\epsilon$, it can be seen from Fig. 2(a) 
that the generation of the
state\thinspace $ {\textstyle{1 \over \sqrt{2}}}
\left( \left| 00\right\rangle +\left| 11\right\rangle \right) $ requires a
pulse of length $\tau _{_{B}}=7.7\times 10^{-15}$ s. By changing the value of
the amplitude $A$, we can modify
the length $\tau _{_{B}}$ of this Bell pulse, i.e. a new $A$ implies a new 
value for $\tau _{_{B}}$: from Fig. 2 we can see that $\tau _{_{B}}$ can be tailored in such 
a way that reliable entangled 
state preparation can be done in the interval $10^{-11}$ s $<\tau _{_{B}}<10^{-15}$ s
\cite{reina1}, which is in agreement
with currently available excitonic dephasing times \cite{Bonadeo1}.

\subsubsection{Three coupled QDs and GHZ states}

We give the procedure for generating the entangled GHZ
states $\left| \Psi _{GHZ}(\varphi)\right\rangle =
{\textstyle{1 \over \sqrt{2}}} (\left| 000\right\rangle
+e^{i\varphi }\left| 111\right\rangle ),$ for arbitrary values of
$\varphi, $ in the proposed system of $3$ coupled QDs. Without loss of generality, 
we
consider the $J=\frac{3}{2}-$subspace as the only one optically active (the other 
two $J=1/2$ subspaces
remain optically dark). We work in the basis set $|J=3/2,M\rangle$,
$\{|0\rangle=|3/2,-3/2\rangle$, $|1\rangle=|3/2,-1/2\rangle$,
$|2\rangle=|3/2,1/2\rangle$, $|3\rangle=|3/2,3/2\rangle\}$,  where $|0\rangle$ 
is the vacuum state, 
$|1\rangle$ is the single-exciton state, $|2\rangle$ is the biexciton state and 
$|3\rangle$ is the triexciton
state. In the absence of light, the energy levels of the system are given by 
$E_{0}\equiv E(3/2,-3/2)=
{\textstyle{3 \over 2}}
(W-\Delta _{\omega })$, $E_{1}\equiv E(3/2,-1/2)=
{\textstyle{1 \over 2}}
(7W-\Delta _{\omega })$, $E_{2}\equiv E(3/2,1/2)=
{\textstyle{1 \over 2}}
(7W+\Delta _{\omega })$, and $E_{3}\equiv E(3/2,3/2)=
{\textstyle{3 \over 2}}
(W+\Delta _{\omega })$. Next we consider, at resonance, the effect of the
pulse of light $\xi (t)$  over this system of 3 QDs: we get the new
eigenenergies $
E_{0,1}={\textstyle{5 \over 2}}
W+\left| A\right| \pm \sqrt{\left( W+\left| A\right| \right) ^{2}+3\left|
A\right| ^{2}},$ and $E_{2,3}={\textstyle{5 \over 2}}
W-\left| A\right| \pm \sqrt{\left( W-\left| A\right| \right) ^{2}+3\left|
A\right| ^{2}}.$

\begin{figure}
%\psdraft
\centerline{\epsfig{file=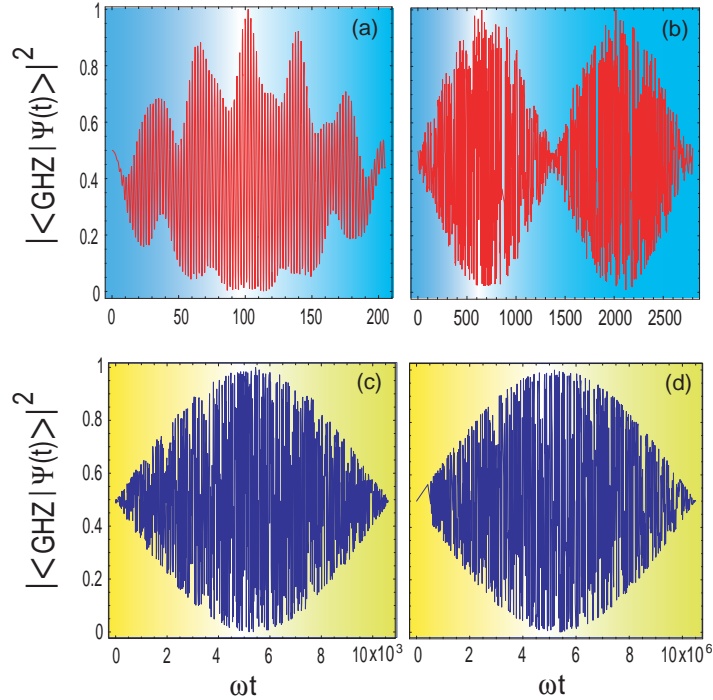,width=9.5cm,height=9.5cm}}
%\vspace{0.2cm}
\caption{\small Generation of the GHZ state ${\textstyle{1 \over \sqrt{2}}}
(\left| 000\right\rangle +\left| 111\right\rangle )$.\ These pulses correspond
to the realization of the Hadamard gate followed by two quantum CNOT
gates. $W=0.1$,$\;\varphi =0,\;$ and 
(a)$\; A={\textstyle {1 \over 25}}$, (b)$\; A={\textstyle {1 \over
50}}$, (c)$\;A=10^{-2}$, and (d)$\;A=10^{-3}$.}
\label{figure3}
\end{figure}

Starting with a zero-exciton state as the
initial state, i.e. $\left| \Psi _{0}\right\rangle =\left| 0\right\rangle$, 
we calculate the 
probability density $\wp (GHZ)\;$of finding the entangled 
$\left| \Psi _{GHZ}(\varphi)\right\rangle$ state
between vacuum and triexciton states as
\begin{equation}
\wp (GHZ)=
{\textstyle{1 \over 2}}
\left| 
\mathop{\displaystyle\sum}
_{k}C_{k}\left(
A_{k1}+e^{i\varphi }A_{k4}\right) e^{-iE_{k}t}\right| ^{2}.  
\end{equation}
In Figure 3 the selective pulses used to generate the GHZ state $
{\textstyle{1 \over \sqrt{2}}}
\left( \left| 000\right\rangle +\left| 111\right\rangle \right) 
$ ($\varphi=0,2\pi$) 
are shown: it can be seen from Fig. 3(a) that for
a band gap $\epsilon =2.8\;$eV  (resonant
optical frequency $\omega =4.3\times 10^{15}$ s$^{-1}$),
$W=0.1$ $\epsilon\;$and$\;A={\textstyle{\epsilon \over 25}}$, 
a pulse of length
$\tau _{_{GHZ}}=1.3\times 10^{-14}$ s is required. We explore 
several different ranges for the $\tau
_{_{GHZ}}-$pulses required in the generation of these GHZ
states. For fixed $W,$ the time 
$\tau _{_{GHZ}}$ increases with decreasing incident field
strength $A.$ In contrast, for fixed $A,$ the length $
\tau _{_{GHZ}}$ decreases with decreasing interdot interaction
strength \cite{reina1}. It is worth noting that after the
preparation step, which is determined by the length of the pulses $
\tau _{_{Bell}}$ and $
\tau _{_{GHZ}}$, the F\"orster interaction parameter $W$, and the field 
strength $A$, the system will evolve under the
action of the Hamiltonian (1) with $\Delta_\omega=A=0$: each one of 
the maximally entangled states discussed
here are eigenstates of this remaining Hamiltonian.

The above results are not restricted to ZnSe-based QDs: 
by employing semiconductors of different bandgap $\epsilon $ (e.g., GaAs,
organic-inorganic systems), other
regions of parameter space can be explored. We have studied the time
evolution of the system of QDs for several different values of the phase $
\varphi $. These give similar qualitative results to the ones discussed 
previously. Next, we show how the density 
matrix formalism can be 
used in an equivalent manner in order to produce 
the excitonic entangled states described before.

\subsection{Pseudo-spin operators and the density matrix}

In  this section we consider a 
rectangular radiation pulse, starting at time
$t=0$ with central frequency $\omega$, given by
$\xi(t)={\it A} {\rm cos}(\omega t)$. The time evolution of any initial state under the
action
of the Hamiltonian (1) is
easily performed by means of the pseudo $\frac {1}{2}-$spin operator 
formalism\cite{wokaun,vega}. Single transition operators are
defined by
\begin{equation}
\langle i|J_x^{r-s}|j\rangle =\frac {1}{2} (\delta_{ir}\delta_{js}+
\delta_{is}\delta_{jr}),\ \
\langle i|J_y^{r-s}|j\rangle =\frac {i}{2} (-\delta_{ir}\delta_{js}+
\delta_{is}\delta_{jr}),\ \
\langle i|J_z^{r-s}|j\rangle =\frac {1}{2} (\delta_{ir}\delta_{jr}-
\delta_{is}\delta_{js})
\end{equation}
where $r$-$s$ denotes the transition between states $|r\rangle$ and
$|s\rangle$
within a  given $J$ subspace. The
three operators belonging to one particular transition $r$-$s$ obey standard
angular momentum commutation relationships
$\left [ J_{\alpha}^{r-s},J_{\beta}^{r-s} \right ]=iJ_{\gamma}^{r-s}$, 
where $(\alpha,\beta,\gamma)$ represents a cyclic permutation of
$(x,y,z)$ (operators belonging to non-connected transitions 
commute:
$\left [ J_{\alpha}^{r-s},J_{\beta}^{t-u} \right ]=0$
with $\alpha, \beta=x,y$ or $z$). In this case, the Hamiltonian in the rotating frame  
($\epsilon\gg W$) becomes\footnote{The Hamiltonian (6) differs from the one given in Eq. (1) by a sign 
because of the choice of the sign for the interdot interaction $W$.} 
\begin{equation}
H_{_\Lambda}=\Delta _\omega J_z - {\textstyle{1 \over 2}}A (J_+ + J_-) -W(J^2-J_z^2).
\end{equation}
We now
give the expressions for the
density  matrix associated with the $N=2,3$ QD systems and show that the 
Hamiltonian (6) leads to the generation of the 
entangled states $\left| \Psi _{Bell}(\varphi)\right\rangle,$ and $\left|\Psi _{GHZ}(\varphi)\right\rangle.$

\subsubsection{Bell states} 
Here we describe the
light excitation procedure to obtain the Bell-type states $|\Psi_{Bell}(\varphi)\rangle$.
To find the analytical solution of the dynamical
equation
governing the system's matrix density, we start with the initial condition 
representing
the vacuum of excitons: only the $J=1$ subspace is
optically active (the $J=0$ subspace remains dark). 
Choosing the basis of eigenstates of $J^2$ and $J_z$ as in Section 2.1.1, 
the rotating frame Hamiltonian and initial
density  matrix can be expressed in terms of pseudo-spin operators as
follows
\begin{eqnarray}
\nonumber
\hspace{-1.0cm}
H_{_{\Lambda}}=-2\Delta _{\omega} J_z^{0-2}+{\textstyle{2W \over 3}}(J_z^{0-1}-J_z^{1-2})-
\sqrt{2} A (J_x^{0-1}+J_x^{1-2})\ \ ,\nonumber 
\hspace{-2.70cm}
\\ 
\rho(0)={\textstyle{1 \over 3}}I+{\textstyle{2 \over 3}}(J_z^{0-1}+J_z^{0-2})\ \ .
\end{eqnarray}
Here $I$ denotes the identity matrix in the subspace $J=1$.
In the absence of light, the energy levels of the system are given as in Section 
2.1.1 (with accuracy of a sign).
Consider the action of
a  pulse of light at resonance and amplitude $A\ll W$. 
Assuming that the
decoherence processes are negligibly small over the time scale of the
evolution (see later), the density matrix at time $t$ becomes
\begin{equation}
\rho (t)={\textstyle{1 \over 3}}I+\left [ {\rm cos}(\omega_2t)+{\textstyle{1 \over 3}}\right ]
J_z^{0-1}+
\left [ {\rm cos}(\omega_2t)-{\textstyle{1 \over 3}}\right ] J_z^{1-2}
-{\rm sin}(\omega_2t)J_y^{0-2} \ \ ,
\end{equation}
which exhibits the generation of coherence between vacuum and
biexciton
states through the operator $J_y^{0-2}$, which oscillates at
frequency $\omega_2=A^2/W$.
\par The state $|\Psi_{Bell}(\varphi)\rangle$ has a
corresponding density matrix
$\rho_{Bell}=I/3+J_z^{0-1}/3-J_z^{1-2}/3+{\rm cos}(\varphi)J_x^{0-2}-
{\rm sin}(\varphi)J_y^{0-2}$.
Comparing this last equation with Eq. (8), we see that the system's quantum
state
at time $\tau_B=\pi W/2A^2$ corresponds to the maximally entangled Bell state
$|\Psi_{Bell}(\pi/2)\rangle$. The time evolutions of populations and 
coherences for an initial vacuum state are plotted in Fig. 4. The evolution
of populations of the vacuum $\rho_{00}$ and the biexciton $\rho_{22}$ 
states are shown in Fig. 4(a).
Clearly the approximate analytic calculation given here describes the
system's evolution very well when compared with the exact
numerical  solution (Fig. 4(a)). Figure 4(b) shows the 
overlap, $O(t)=Tr \left [ \rho_{Bell}\rho(t) \right ]$, 
between the maximally entangled Bell state and the one
obtained by applying a rectangular pulse of light at resonance.
The thick solid line (Fig. 4(b)) describes $O(t)$ with a maximally entangled
Bell
state in the rotating frame, while the thin solid line (Fig. 4(b)) represents
the
overlap with a Bell state transformed to the laboratory frame: obviously the
rotating frame
case corresponds to the amplitude evolution of the  laboratory frame signal.
The dashed line illustrates the approximate solution overlap in
the rotating frame.  The approximate solution works very well,
supporting the idea that a selective Bell pulse of length
$\tau_B=\pi W/2A^2$ can be used to create the Bell state 
$|\Psi_{Bell}(\pi/2)\rangle$ in
the system of two coupled QDs. 
The same conclusion can also be drawn from the time evolution
of the overlap between the exact Bell-state density matrix and the one
obtained
directly from the numerical calculation \cite{reina1}(b). Therefore, 
the existence of a selective Bell pulse is numerically confirmed.

\begin{figure}
%\psdraft
\centerline{\epsfig{file=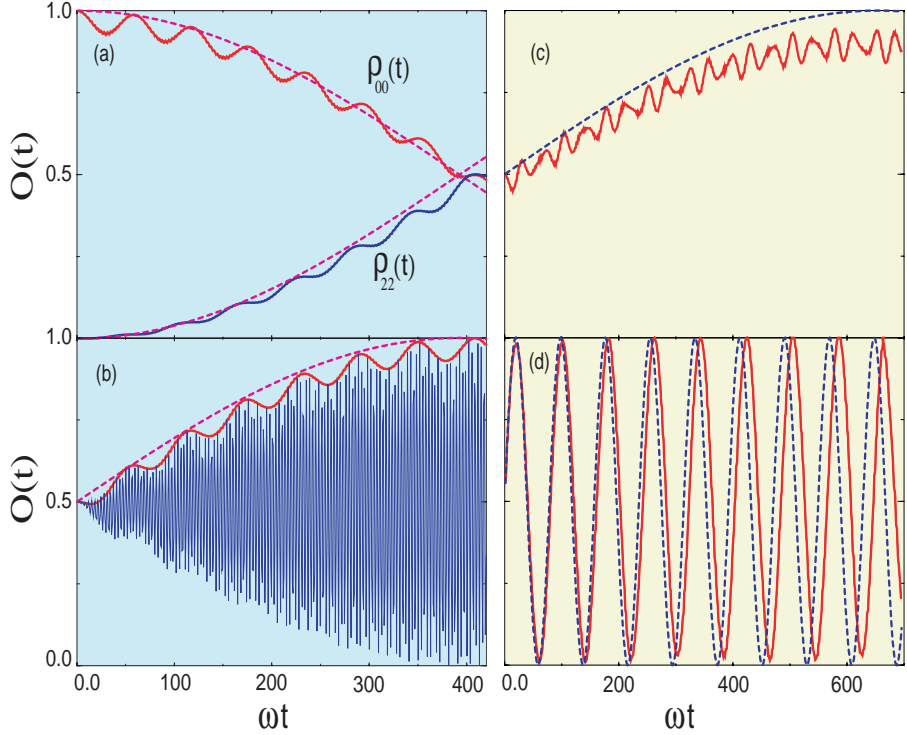,width=12cm,height=10cm}}
\caption{\small (a) Population of the vacuum state $\rho_{00}$ and
biexciton state $\rho_{22}$ in two coupled QDs, as a function of time.
(b) Time-evolution of overlap with maximally entangled Bell state.
$\epsilon=1$, 
$W=\epsilon/10$ and $A=W/5$. Blue (thin solid) line shows exact numerical result in
the laboratory frame. Red (thick solid) line in (b) represents the exact
numerical solution in the rotating frame. Pink (dashed) line shows approximate
analytical result.
(c) Time-evolution of overlap with maximally
entangled $GHZ$ states  $\left| \Psi _{GHZ}(\varphi)\right\rangle$, and (d)  $\left| \Psi
_{GHZ}(\varphi)\right\rangle_2$,  under the
action of a rectangular pulse of light at resonance. $\epsilon=1$, 
$W=\epsilon/10$ and $A=2W/5$. Red (solid) line represents exact numerical
solution. Blue (dashed) line shows approximate analytical result.}
\label{figure4}
\end{figure}

\subsubsection{GHZ states}
Next, consider three
quantum dots of equal size placed at the corners of an equilateral triangle, 
as in Section 2.1.2,
with the $J=\frac{3}{2}$ subspace being the only one optically active subspace, and with the 
same basis
set of Section 2.1.2. In terms of pseudo-spin operators, the rotating frame Hamiltonian, including the 
radiation term, is now given by
\begin{equation}
H_{_\Lambda}=-\Delta_\omega (3J_z^{0-3}+J_z^{1-2})+2W(J_z^{0-1}-2J_z^{2-3})- 
A \left [ \sqrt{3} (J_x^{0-1}+J_x^{2-3})+2J_x^{1-2} \right]\ \ .
\end{equation}
In terms of its associated density matrix, the entangled state $\left| 
\Psi _{GHZ}(\varphi)\right\rangle$
between vacuum
and triexciton states is given by 
$\rho_{GHZ}=I/4+J_z^{0-1}/2
-J_z^{2-3}/2+{\rm cos}(\varphi)J_x^{0-3}+{\rm sin}(\varphi)J_y^{0-3}$, where $I$
denotes the identity matrix in the $J=\frac{3}{2}$ subspace.
This state can be generated after an appropriate $\frac{\pi}{2}-$pulse: starting
with a zero-exciton state $|0\rangle$, at resonance, and using the
properties of
pseudo-spin operators, the evolved state under the action of
Hamiltonian Eq. (9) can
be obtained in a straighforward way in the limit $A/W\ll 1$ \cite{reina1}(b):
\begin{eqnarray}
\rho(t)={\textstyle{1 \over 4}}I+\left [ {\rm cos}(\omega_3t)+{\textstyle{1 \over 2}}\right ]
J_z^{0-1}+
{\rm cos}(\omega_3t)J_z^{1-2}+
\left [ {\rm cos}(\omega_3t)-{\textstyle{1 \over 2}}\right ] J_z^{2-3}
+{\rm sin}(\omega_3 t) J_y^{0-3} \ \ ,
\end{eqnarray}
with $\omega_3=d_--d_++A$ and 
$d_\pm=W \left [ 1\pm\frac {A}{W}+(\frac {A}{W})^2 \right] ^{1/2}$.
Clearly $|\Psi_{GHZ}(\pi/2)\rangle$ can be generated with 
a $\frac{\pi}{2}-$pulse of length $\tau_{GHZ}=4\pi W^2/3A^3$.
In Fig. 4(c) we show the overlap between the exact density matrix
and that corresponding to state $\left| \Psi _{GHZ}(\varphi)\right\rangle$. The dashed line
shows the overlap using our approximate density matrix, Eq. (10). 

We also give the scheme 
for generating the entangled
state between a single exciton $|1\rangle$ and the  biexciton $|2\rangle$, 
$\left| \Psi
_{GHZ}(\varphi)\right\rangle_2=\frac {1}{\sqrt{2}}(|1\rangle+e^{i\varphi}|2\rangle)$. 
In order to generate $\left| \Psi
_{GHZ}(\varphi)\right\rangle_2$, we take the single exciton state
$|1\rangle$ as the initial condition. Evolution of this new initial state
under $H_{_\Lambda}$ (Eq. (9)) with $\Delta_\omega=0$
generates a new density matrix $\rho (t)$ which can be used to show that a
pulse of duration
$\tau_{GHZ}^{\prime}=\pi/4A$, generates the state $\left| \Psi
_{GHZ}(\pi/2)\right\rangle_2$. Figure 4(d) shows the overlap between $\rho(t)$
and $\rho_{GHZ_2}$ \cite{reina1}(b). We emphasize that the two maximally entangled 
GHZ states considered above have very
different frequencies. This feature
should enable each of
these maximally entangled $GHZ$ states to be manipulated separately in
actual experiments, even if the initial state is mixed.

From the results above,
it follows that in order to generate maximally entangled exciton states,
$\frac{\pi}{2}-$pulses with sub-picosecond duration 
should be used. A surprising conclusion of our results is that entangled-state preparation is
facilitated by {\em weak} light fields (i.e. $A\ll W$): strong fields cause
excessive oscillatory behavior in the density matrix. 
The relevant
experimental conditions as well as the required coherent control to realize the
above combinations of parameters, are compatible with those demonstrated in
Refs. \cite{Bonadeo2,Bonadeo1,axt}: we expect that the experimental generation 
of the Bell and GHZ states discussed here should be possible with these ultrafast
semiconductor optical techniques. Here it is important to highlight that the
corresponding increase in the effective gap will yield a larger exciton binding
energy: typical decoherence  mechanisms (e.g., acoustic phonon scattering) will
hence become less effective. The generation of maximally entangled states in this
proposal has considered the experimental situation of {\it global} excitation
pulses, i.e. pulses acting simultaneously on the entire QD system. However, by
using near-field optical spectroscopy \cite{ChavezPirson}, individual QDs from an
ensemble can be addressed by using local pulses, a feature that can be
exploited to generate entangled states with different symmetries, such as the
antisymmetric state
${\textstyle{1 \over \sqrt{2}}} (\left| 01\right\rangle -\left|
10\right\rangle).\;$ Hence, we should be able to generate the so-called Bell
basis of four mutually orthogonal states for the 2 qubits, all of
which are maximally entangled, i.e. the set of states
${\textstyle{1 \over \sqrt{2}}}\{(\left| 00\right\rangle +\left|
11\right\rangle),\; (\left| 00\right\rangle -\left| 11\right\rangle),\; (\left|
01\right\rangle +\left| 10\right\rangle),\; (\left|
01\right\rangle -\left| 10\right\rangle)\} $.$\;$ From a general point of
view, this basis is of fundamental relevance for quantum
information processing. 

In summary, we have shown how maximally entangled Bell and GHZ states 
can be generated using the optically driven resonant transfer
of excitons between quantum dots. Selective Bell and GHZ pulses have been 
identified
by an approximate, yet accurate, analytical approach which should prove a
useful
tool when designing experiments. Exact numerical calculations 
confirm the existence of such
$\varphi-$pulses for the generation of maximally entangled states in coupled
dot systems. 

\section{Decoherence mechanisms}

Here we analize the reliability of the preparation of entangled states when
decoherence mechanisms are taken into account during the generation step. Exciton
decoherence in semiconductor QDs is dominated by acoustic phonon scattering at
low temperatures \cite{taka}. Hence, we consider the acoustic phonon dephasing
mechanism

\begin{eqnarray}
H_{env}=\sum_{\vec k}
{\omega_{\vec k}a_{\vec k}^{\dag} a_{\vec k}}+\sum_{\vec k}
{g_{\vec k}J_z(a_{\vec k}^{\dag}+a_{\vec k})},
\end{eqnarray}
where $a_{\vec k}^{\dag}$
($a_{\vec k}$) is the creation (annihilation) operator of the acoustic phonon with
wavevector ${\vec k}$, as the main factor responsible for decoherence effects in the
generation of the maximally entangled exciton states analized
before \cite{ferney}. The new time evolution to be analyzed
is modelled by the Hamiltonian
$H'=H_{_\Lambda}+H_{env}$. 
\begin{figure}
%\psdraft
\centerline{\epsfig{file=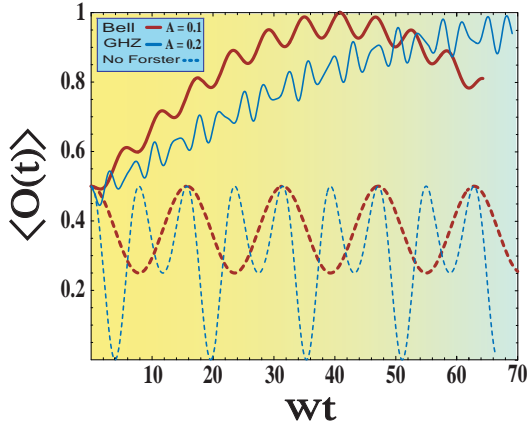,width=7.0cm}}
%\vspace{0.2cm}
\caption{\small Maximally entangled exciton states generation in the zero 
decoherence limit. Thick (red)
lines represent the Bell-state overlap with $A=0.1$: solid, Forster 
term included; dotted, 
Forster term not included. Thin (blue) lines represent the GHZ-state overlap with 
$A=0.2$ and same meaning for solid and dotted lines.}
\label{figure5}
\end{figure}
Here we consider pure decoherence effects
that do not involve energy relaxation of excitons 
(these effects will be addressed elsewhere \cite{reina4}). The exact kinetic 
equations for the system of QDs can be obtained by 
applying the method of operator-equation hierarchy developed for
Dicke systems in \cite{bogo}. Following the standard procedure, by assuming
a very short correlation time for exciton operators, the exact
hierarchy
of equations transforms into a Markovian master equation.  The initial
condition is represented by the density matrix
$\rho(0)=|0\rangle\langle 0|
\rho_{Ph}(T)$,
exciton vacuum and the equilibrium phonon reservoir
at temperature $T$.
At resonance ($\Delta_\omega=0$) the dynamical equation for the
expectation value of exciton operators is given by
\begin{equation}
\frac {\partial \langle J_{\alpha}^{r-s}\rangle}{\partial t}=
-iW\langle [J_{\alpha}^{r-s},J_z^2]\rangle
-iA\langle [J_{\alpha}^{r-s},J^++J^-]\rangle-
\Gamma(2\langle [J_{\alpha}^{r-s},J_z]J_z\rangle - 
\langle [J_{\alpha}^{r-s},J_z^2] \rangle),
\end{equation}
where 
$\Gamma=\int {d\omega' \omega'^n
e^{-\omega'/\omega_c}(1+2N(\omega',T))}$ is the decoherence rate
with $n$ depending on the dimensionality of the
phonon field, $\omega_c$ is a cut-off frequency (typically the Debye
frequency) and
$N(\omega',T)$ is the phonon Bose-Einstein occupation factor.
It is a well known fact that
very narrow linewidth of the photoluminescence signal of a {\it single} 
QD does exist due to the
elimination of inhomogeneous broadening effects. Consequently, the 
decoherence rate $\Gamma$ in this 
analysis should be associated with just homogeneous
broadening effects. At low temperature the main decoherence mechanism is 
indeed acoustic phonon 
scattering processes. The decoherence parameter
$\Gamma$ is temperature dependent and it amounts to 20-50 $\mu$eV for 
typical III-V semiconductor QDs
in a temperature range from 10 K to 30 K \cite{taka}. We consider typical 
values for $\Gamma$
which can represent real situations for QDs at low temperatures together 
with the experimental conditions
$\epsilon=\omega=1$ and the F\"{o}rster term $W=0.1$ $\epsilon$. 
Laser strengths and decoherence rates are expressed in units of $W$. 
The coupled differential 
linear equations for the time dependent pseudo-spin expectation values are 
solved and the results 
are given in terms of
the time dependent overlaps
$O_B(t)=Tr\{\rho_{Bell}\rho(t)\}$ and $O_{GHZ}(t)=Tr\{\rho_{GHZ}\rho(t)\}$ \cite{ferney}.

\begin{figure}
%\psdraft
\centerline{\epsfig{file=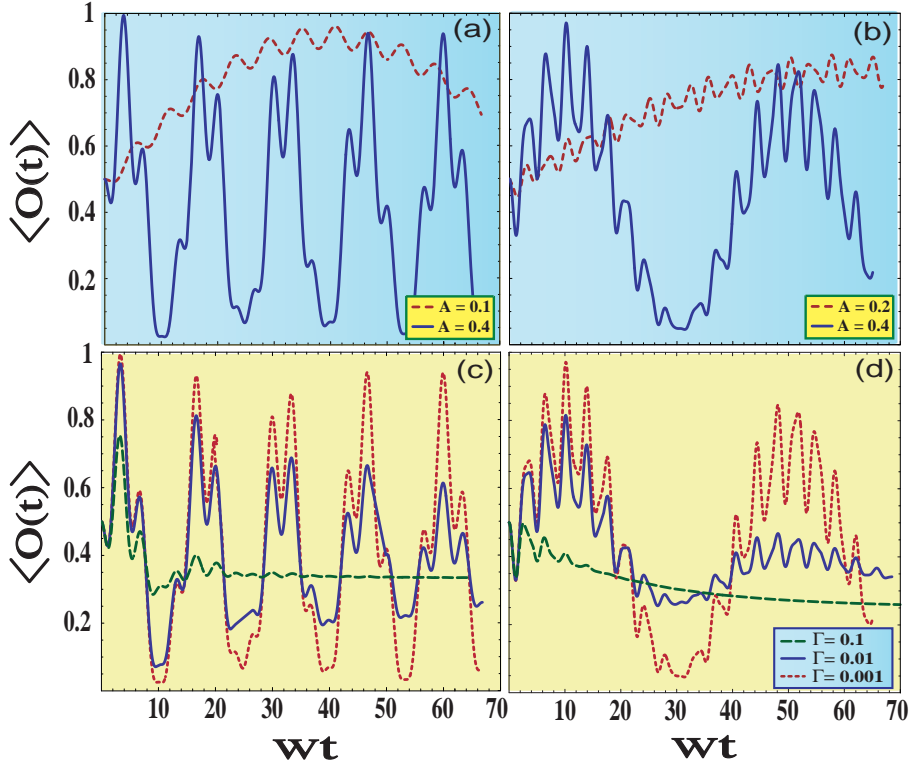,width=12.0cm}}
%\vspace{0.2cm}
\caption{\small Maximally entangled exciton states generation in the presence of decoherence:
 (a)  $\langle O_B(t)\rangle$ for $A=0.1$, red (dotted) line and
$A=0.4$, blue (solid) line. (b) $\langle O_G(t) \rangle$ for $A=0.2$, red (dotted) line and
$A=0.4$, blue (solid) line. In plots (a) and (b) $\Gamma$ is kept fixed: $\Gamma=0.001$. 
In figures (c) and (d) $A=0.4$ is kept fixed whereas $\Gamma$ is varied:
$\Gamma=0.001$, red (dotted) line,
$\Gamma=0.01$, blue (solid) line and $\Gamma=0.1$, green (dashed) line. These curves correspond to (c)
$\langle O_B(t)\rangle$ and (d) $\langle O_G(t) \rangle$.}
\label{figure6}
\end{figure}

Figure 5 shows
the evolution of the
overlaps $O_B(t)$ and $O_{GHZ}(t)$ in the
limit of very weak light excitation and zero decoherence.
It can be seen that no maximally entangled exciton states generation 
is possible if the
F\"orster interaction is turned off. This implies that
efficient exciton entangled states generation should be helped by compact
QD systems where the F\"orster term can take a significant value, as we 
discussed in the above section.
Figures 6(a) and 6(c) show the case of Bell-state generation ($N=2$ QDs) 
in the presence of noise. In Fig. 6(a) the
decoherence rate $\Gamma=0.001$, and the laser intensities are $A=0.1$, 
and $A=0.4$. 
It is shown that $\tau_B$ is significantly shortened by
applying stronger laser pulses. Therefore, decoherence effects can be 
minimized by using higher excitation levels. 
However, a higher laser
intensity also implies a sharper evolution which therefore requires a 
very precise pulse length. Figure 6(c) shows
temperature-dependent results for $\Gamma=0.001$, 0.01 and 0.1, when 
$A=0.4$ is kept fixed. We can see that at high
temperatures ($\Gamma=0.1$) no maximally entangled states generation 
is possible. However, it can be estimated that
$\Gamma$ values between $0.001-0.01$ are typical in the temperature 
range from $10$ K to $50$ K: in this
parameter window successful generation of Bell 
states can be produced \cite{ferney}, as shown in Fig. 6(c).
Figures 6(b) and 6(d) show the case of GHZ states generation ($N=3$ QDs). 
As above, $\tau_{GHZ}$ is shortened by using
higher laser excitation levels, as can be seen from Fig. 6(b) for $\Gamma=0.001$. 
Figure 6(d) shows the temperature
effects through the variation of $\Gamma$ for $A=0.4$. We see that similar 
decoherence rates yield  a more
dramatic reduction of the coherence in the GHZ case than in the Bell case. 
However, as for Bell
generation, a parameter window does exist where the 
generation of such entangled states are
feasible \cite{ferney}. It is worth noting the different scaling behaviour 
of the generation
frequency of these entangled states at very low temperature, i.e. 
vanishing $\Gamma$ and
very low laser excitation. While 
selective $\frac{\pi}{2}$ laser pulse length for the Bell case
scales like $W/A^2$, selective $\frac{\pi}{2}$ pulse length for the
GHZ case scales like $W^2/A^3$. This property of $\frac{\pi}{2}$ pulses to 
generate maximally entangled
exciton states was demonstrated analytically in the above 
section and is
verified in the present section by looking at the numerical results presented 
in Figure 6. 

In summary, decoherence effects can be minimized in the generation of maximally 
entangled states by
applying stronger laser pulses and working at low temperatures where acoustic 
phonon scattering is the
main decoherence mechanism.
Since we have shown that the generation of maximally entangled exciton states is
preserved over a reasonable parameter window even in the presence of decoherence 
mechanisms, 
we stress that this optical
generation could be exploited in
solid state devices to perform quantum protocols, such as the teleportation 
of an excitonic 
state in a coupled QD system \cite{reina3}, as we show next.

\section{Quantum teleportation of excitonic states }

Here we propose a
practical scheme capable of demonstrating quantum teleportation which exploits 
currently
available ultrafast spectroscopy techniques in order to prepare and
manipulate entangled states of excitons in coupled QDs \cite{reina3}. Since the 
original idea of quantum 
teleportation considered in 1993 by
Bennett et al. \cite{Bennett1}, great efforts have been made to
realize the physical implementation of teleportation devices \cite
{Teleportation}. The general scheme of teleportation \cite{Bennett1}, which
is based on Einstein-Podolsky-Rosen (EPR) pairs \cite{Einstein} and Bell
measurements \cite{Bell} using classical and purely nonclassical
correlations, enables the transportation of an arbitrary quantum state from
one location to another without knowledge or movement of the state itself
through space. This process has been explored from various points of view 
\cite{Teleportation}; however none of the experimental set-ups to date have
considered a solid-state approach, despite the recent advances in
semiconductor nanostructure fabrication and measurement \cite
{Neil,Bonadeo2,Bonadeo1,ChavezPirson}. Reference \cite{Bonadeo1}, for
example, demonstrates the remarkable degree of control which is now possible
over quantum states of individual quantum dots (QDs) using ultra-fast
spectroscopy. The possibility therefore exists to use optically-driven QDs
as ``quantum memory'' elements in quantum computation operations, via a
precise and controlled excitation of the system. 

In order to implement the quantum operations for the description of the
teleportation scheme proposed here, we employ two elements: the Hadamard
transformation and the quantum
controlled-NOT\ gate (C-NOT gate). In the orthonormal computation basis of single qubits $
\left\{ \left| 0\right\rangle ,\left| 1\right\rangle \right\} $, the C-NOT
gate acts on two qubits $\left| \varphi _{i}\right\rangle $ and $\left|
\varphi _{j}\right\rangle $ simultaneously as follows: C-NOT$_{ij}(\left|
\varphi _{i}\right\rangle \left| \varphi _{j}\right\rangle )\mapsto \left|
\varphi _{i}\right\rangle \left| \varphi _{i}\oplus \varphi
_{j}\right\rangle $. Here $\oplus $ denotes addition modulo 2. The
indices $i$ and $j$ refer to the control bit and the target bit
respectively (see Fig. 7). The Hadamard gate $U_H$ acts only on single qubits by
performing the rotations $U_H(\left| 0\right\rangle )\mapsto \frac{1}{\sqrt{2}}
\left( \left| 0\right\rangle +\left| 1\right\rangle \right)$ and $U_H(\left|
1\right\rangle )\mapsto \frac{1}{\sqrt{2}}\left( \left| 0\right\rangle
-\left| 1\right\rangle \right) $. The above unitary transformations can be
written as

%\newpage
\begin{equation}
U_{H}=
{\textstyle {1 \over \sqrt{2}}}\left (\matrix{ 1 & ~~1\cr 1 & -1\cr}\right),\ \
C-NOT=\left(\matrix{1&0&0&0\cr 0&1&0&0\cr 0&0&0&1\cr 0&0&1&0\cr}
          \right) \ \ ,
\end{equation}\\
and represented in the language of quantum circuits as in Figure 7.
\begin{figure}
%\psdraft
\centerline{\epsfig{file=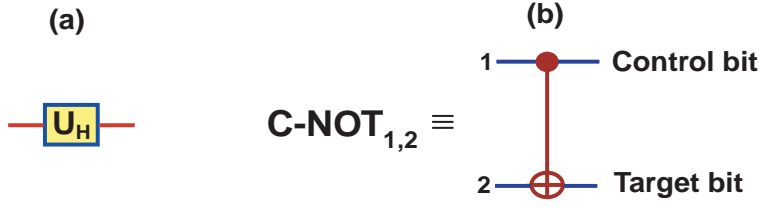,width=10cm}}
\caption{\small Schematic representation of (a) The Hadamard gate, and (b) The controlled-NOT gate.}
\label{figure7}
\end{figure}
We also introduce a pure state $\left|
\Psi \right\rangle $ in this Hilbert space given by $\left| \Psi
\right\rangle =\alpha \left| 0\right\rangle +\beta \left| 1\right\rangle $
with\ $\left| \alpha \right| ^{2}+\left| \beta \right| ^{2}=1$, where $
\alpha \;$and $\beta \;$are $\mathop{\rm complex}$\ numbers. As discussed 
in the above section,
$\left| 0\right\rangle$ represents the vacuum state for excitons while $\left| 1\right\rangle$ 
represents
a single exciton. Following Ref. \cite{Brassard}, in Figure 8 we show the general computational 
approach discussed in this section. 
As usual, we refer to two
parties, Alice and Bob. Alice wants to teleport an arbitrary, unknown qubit
state $\left| \Psi \right\rangle $ to Bob. Figure 9 shows the specific
realization we are proposing using optically controlled quantum dots with QD 
$a$ initially containing $\left| \Psi \right\rangle $. Alice prepares two
qubits (QDs $b$ and $c$) in the state $\left| 0\right\rangle $ and then
gives the state $\left| \Psi 00\right\rangle$ as the {\it input} to the
system. By performing the series of transformations shown in Fig. 8, Bob
receives as the {\it output} of the circuit the state $
{\textstyle {1 \over \sqrt{2}}}
(\left| 0\right\rangle +\left| 1\right\rangle )_{a}
{\textstyle {1 \over \sqrt{2}}}
(\left| 0\right\rangle +\left| 1\right\rangle )_{b}\left| \Psi \right\rangle
_{c}$ (Fig. 9(d)). In Ref. \cite{reina3} we generalize the teleportation
scheme given in Ref. \cite{Brassard} to the case of an $N$ qubit quantum circuit.
In order to describe the physical implementation of the quantum circuit given in Fig. 8
using coupled quantum dots, we exploit the recent experimental results
involving coherent control of excitons in single quantum dots on the
nanometer and femtosecond scales \cite{Bonadeo1,Bonadeo2}. Consider a system
of three identical and equispaced QDs containing no net charge (Fig. 9(a)),
which are initially prepared in the state $\left| \Psi \right\rangle
_{a}\left| 0\right\rangle _{b}\left| 0\right\rangle _{c}$. As shown in Fig.
9(a), one of these (QD $a$) contains the quantum state $\left| \Psi
\right\rangle _{a}$ that we wish to teleport, while the other two (QDs $b$
and $c$) are initialized in the state $\left| 00\right\rangle _{bc}-$ this
latter state is easy to achieve since it is the ground state. Following this
initialization, we illuminate QDs $b$ and $c$ with the radiation pulse 
$\xi (t)=A\exp (-i\omega t)$ (see Fig. 9(b)). For the case of ZnSe-based QDs, 
the band gap $\epsilon
=2.8$ eV, hence the resonance optical frequency $\omega =4.3\times 10^{15}$ s$^{-1}$. 
For a $0$ or $2\pi-$pulse, the density of probability for finding the QDs $b$ and $c$
in the Bell
state\thinspace $ {\textstyle {1 \over \sqrt{2}}}
\left( \left| 00\right\rangle +\left| 11\right\rangle \right) $ requires the length
$\tau _{_{Bell}}=7.7\times
10^{-15}$ s (see Fig. 2(a)). Hence, this time $\tau _{_{Bell}}$ corresponds 
to the realization of the
first two gates of the circuit in Fig. 8, i.e. the Hadamard
transformation over QD $b$ followed by the C-NOT gate between QDs $b$ and $c$. 
After this, the
information in qubit $c$ is sent to Bob and Alice keeps in her memory the state of QD $b$. 
Next, we need
to perform a  C-NOT operation between QDs $a$ and $b$ and, following that, a Hadamard
transform over the QD $a$: this procedure then leaves the system in the state 
\begin{figure}
%\psdraft
\centerline{\epsfig{file=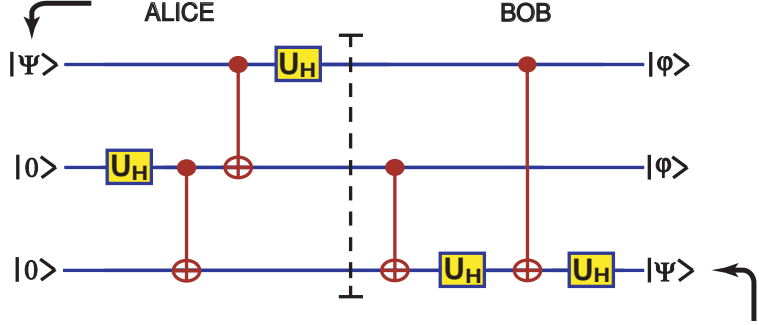,width=10cm}}
\caption{\small Circuit scheme to teleport an unknown quantum state from Alice to
Bob using an arrangement of 3 qubits (coupled quantum
dots).}
\label{figure8}
\end{figure}
\begin{equation}
{\textstyle {1 \over 2}}
\left\{ \left| 00\right\rangle (\alpha \left| 0\right\rangle +\beta \left|
1\right\rangle )+\left| 01\right\rangle (\beta \left| 0\right\rangle +\alpha
\left| 1\right\rangle )+\left| 10\right\rangle (\alpha \left| 0\right\rangle
-\beta \left| 1\right\rangle )+\left| 11\right\rangle (-\beta \left|
0\right\rangle +\alpha \left| 1\right\rangle )\right\}.
\end{equation}
As can be seen from Eq. (14), we are proposing the realization of the Bell 
basis measurement in two steps \cite{Brassard}: 
first, we have rotated from the Bell basis into the computational basis $(\left| 00 \right\rangle $,
$\left| 01 \right\rangle $, $\left| 10 \right\rangle $, $\left| 11
\right\rangle )$, by performing the unitary operations shown before the dashed line 
in Fig. 8. Hence, the second step is to perform a measurement in this computational
basis. At this point, we leave QDs {\it a} and {\it b} in one of the four
states
$\left| 00
\right\rangle $,
$\left| 01 \right\rangle $, $\left| 10 \right\rangle $, $\left| 11
\right\rangle $ (see Fig. 9(c)), which are the four possible measurement results. 
This last step can be experimentally realized by
using {\it near-field optical spectroscopy} \cite{ChavezPirson}.  In this way, 
it is possible to scan,
dot-by-dot, the optical properties of the entire dot ensemble, and particularly, to
measure directly the excitonic photoluminiscence spectrum of dots $a$ and $b$, 
thus completing the Bell 
basis measurement. The result of this measurement provides us with two classical 
bits of information, 
conditional the states measured by nanoprobing
on QDs {\it a} and {\it
b} (see Fig. 9(c)). These classical bits are essential for completing 
the teleportation process: 
rewriting Eq. (14) as
\begin{equation}
\begin{array}{c}
\frac{1}{2}\left\{ \left| 00\right\rangle \left| \Psi \right\rangle
+\left| 01\right\rangle \sigma_{x}\left| \Psi \right\rangle+\right.  
\left. \left| 10\right\rangle \sigma_{z}\left| \Psi \right\rangle+\left|
11\right\rangle (-i\sigma_{y})\left| \Psi \right\rangle\right\}
\end{array}
\end{equation}
we see that if, instead of performing the set of operations shown after
the dashed line in Fig. 8, Bob
performs one of the conditional unitary
operations $ I,\sigma_{x},
\sigma_{z}$, or $-i\sigma_{y}$ over the QD $c$ (depending on the 
measurement results 
or classical signal communicated from Alice to Bob, as shown in Fig. 9(c))\footnote{These unitary transformations, which depend on the
result of Alice's  measurement (subindices of $U$), are the Pauli matrices\ \
$U_{00}\equiv I=\left( 
\begin{array}{ll}
1 & 0 \\ 
0 & 1
\end{array}
\right) $,$\ \ U_{01}\equiv \sigma_x=\left( 
\begin{array}{ll}
0 & 1 \\ 
1 & 0
\end{array}
\right), \ \ U_{10}\equiv \sigma_z=\left(\begin{array}{cc}
-1 & 0 \\ 
~~0 & 1
\end{array}
\right) $,$\ \ U_{11}\equiv -i\sigma_x=\left( 
\begin{array}{cc}
0 & -1 \\ 
1 & ~~0
\end{array}
\right) \ \ . $}  
the teleportation 
process is finished since the excitonic state $\left| \Psi \right\rangle$ 
has been teleported 
from dot $a$ to dot $c$. For this reason only two unitary
exclusive-or transformations are needed in order to teleport the state $
\left| \Psi \right\rangle $.
This final step can be verified by measuring directly the excitonic luminescence 
from dot $c$, which must correspond to the initial state of dot $a$. 
For instance, if the state to be teleported
is $\left| \Psi \right\rangle \equiv \left| 1 \right\rangle$, the final measurement 
of the 
near-field luminiscence spectrum of dot $c$ 
must give an excitonic emission line of the same wavelength and intensity as the 
initial one for dot $a$. 
This measurement process, used for verifying the fidelity of the process, can be 
used if we either 
perform the unitary transformations after Alice's measurement (Fig. 9(c)) or
we realize the complete teleportation circuit shown in Fig. 8, 
leaving the system in the state shown in Fig. 9(d).  
\begin{figure}
%\psdraft
\centerline{\epsfig{file=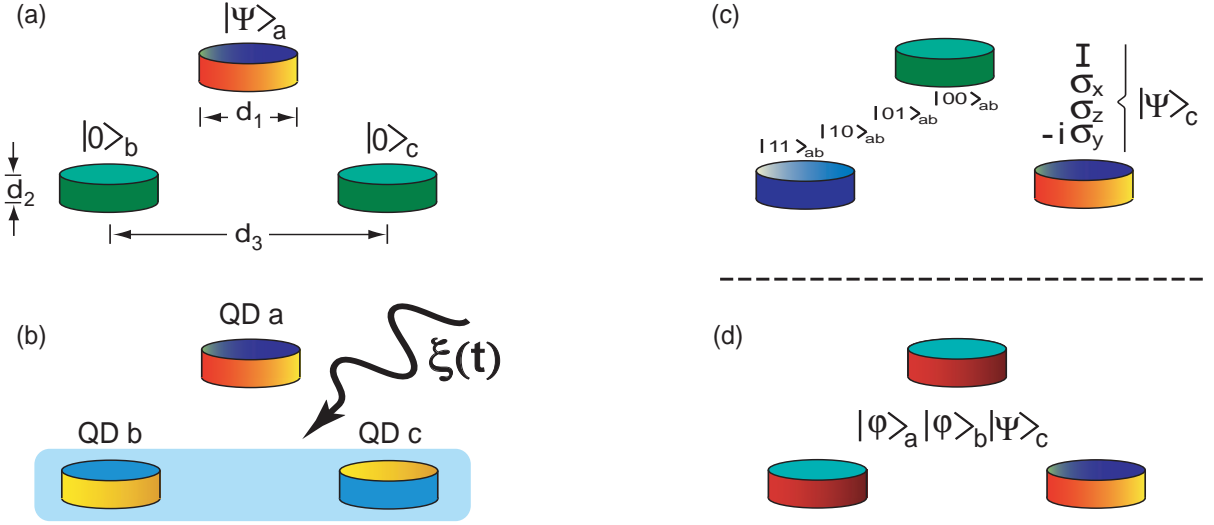,width=16.0cm,height=7cm}}%\bigskip
\caption{\small Practical implementation of teleportation using optically-driven
coupled quantum dots. (a) Initial state of the system. (b) Intermediate
step: radiating the system with the pulse $\xi (t)$. (c) Bell basis measurement and the 
quantum state of the system at the dashed line in Fig. 8. (d) Final state.
Typical values for the dots are diameter $d_{1}=30$ nm, thickness 
$d_{2}=3$ nm and separation $d_{3}=50$ nm.}
\label{figure9}
\end{figure}
As we discussed in Section II, it is possible to excite
and probe just one individual QD with the corresponding dephasing
time $\tau _{d}=4\times 10^{-11}\;$s \cite{Bonadeo1}. Hence we
have the possibility of coherent optical control of the quantum state of a
single dot. Furthermore, this mechanism can be extended to include more than
one excited state: since $
{\displaystyle {\tau _{_{Bell}} \over \tau _{d}}}
\simeq 1.8\times 10^{-4}$,$\;$several thousand unitary operations can in
principle be performed in this system before the excited state of the QD
decoheres. This fact together with the experimental feasibility of applying
the required sequence of laser pulses on the femtosecond time-scale leads us 
to conclude that we do not need to worry
unduly about decoherence ocurring whilst performing the unitary operations
that Bob needs in order to obtain the final states schematically sketched in Figs. 
9(c) and 9(d), 
thereby completing the teleportation process. In the
case of Fig. 3(a), a similar analysis shows that $\tau _{_{GHZ}}=1.3\times 10^{-14}\;$s,\
and hence$\;
{\displaystyle {\tau _{_{GHZ}} \over \tau _{d}}}
\simeq 3.3\times 10^{-4}$: this also makes the 4 qubits circuit given in Ref. \cite{reina3}
experimentally feasible. Although this discussion refers to ZnSe-based QDs,
other regions of parameter space can be explored by employing semiconductors
of different bandgap $\epsilon $. As we will discuss in Section 6, we believe that compact 
hybrid organic-inorganic nanostructures \cite{Agranovich} are very promising  candidates for 
the experimental realization of the setup proposed here. In this case, 
the typical distance between QDs should be of the
same order  as their size: in ZnSe, the Bohr radius of the three dimensional Wannier exciton  
$a_B\approx35$A, hence QDs with radii of about 50A will considerably increase the binding energy
of these excitons. If these dots are placed in an organic matrix 
separated by a distance of the same order, we should be able to
perform the appropriate quantum operations required in the teleportation process 
of the excitonic state $\left| \Psi \right\rangle$. Even though the structures that
we are considering have a dephasing time of order $10^{-11}$ s, QDs with stronger
confinement are expected to have even smaller coupling to phonons giving the
possibility for much longer intrinsic coherence times.

In summary, we have proposed a practical implementation of a semiconductor quantum
teleportation device, exploiting current levels of optical control in
coupled QDs. Furthermore the analysis suggests that several thousand quantum
computation operations may in principle be performed before decoherence
takes place.

\section{Quantum logic with an NMR$-$based nanostructure switch}

Here we propose a novel solid-state based mechanism for quantum computation. The essential 
system is a nuclear
spin$-\frac{1}{2}$ impurity placed at the center of a 2 electrons QD in the presence of an external
perpendicular magnetic field $B$. These electrons undergo abrupt ground-state
transitions as the $B$-field is changed. The different ground states have  very 
different charge distributions and hence
different hyperfine interaction with the nucleus. Thus, by changing $B$ we can  
change the hyperfine coupling and hence tune the
nuclear resonance frequency. This allows one to  effectively select out one such 
dot from an array, and the same mechanism may
also allow an  electron-mediated interaction between nuclei in different dots. 
The proposal is motivated by recent experimental results
which demonstrated the optical detection of an NMR signal in both single
QDs \cite{nmrgammon} and doped bulk semiconductors \cite{kik}. Hence the underlying 
nuclear spins in the QDs can indeed
be controlled with optical techniques, via the electron-nucleus coupling. In addition,
the experimental results of Ashoori et al. \cite{ashoori} and others, have
demonstrated that few electron (i.e. $N \geq 2$) dots can be prepared, and
their magic number transitions measured as a function of magnetic field.
The requirements for the present proposal are therefore compatible
with current experimental capabilities and 
the complications associated with voltage gates or electron transport of other known proposals 
(Refs. \cite{kane,privman}) are avoided by providing an 
{\it all-optical} system.

\subsection{The Model}

As we mentioned briefly before, our model considers an array of silicon-based 
$N-$electron QDs in which impurity
atoms (nuclear spin $\frac{1}{2}$) are placed at the center of each QD (see discussion below). 
Ordinary
silicon ($^{28}$Si) has zero nuclear spin, hence it is possible
to construct the QDs such that no nuclear spins are present other than that
carried by the impurity nuclei, say $^{13}$C. Since carbon is an isoelectronic impurity
in silicon, no Coulomb field is generated by this impurity. Hence
the electronic structure of the bare QDs is essentially unperturbed by the
presence of the carbon atom. 

Suppose the quantum dots are quasi two-dimensional (2D),
contain $N=2$ electrons, and are under the action of the $B$-field. The
lateral confining potential in such quasi-2D QDs is typically parabolic to
a good approximation: the electrons, with effective mass $m^{\ast }$, are
confined by the harmonic potential
$\frac{1}{2}m^{\ast}\omega _{0,i}^{2}
\mbox{$r$}^{2}$, where $\omega _{0,i}$ is in general different for each dot (see Fig. 10). 
We consider two configurations in which all of the electrons in the QDs are confined  
to the (a) $z=0,d,2d,...$ planes (Fig. 10(a)) and
(b) $z=0$ plane (Fig. 10(c)).  The latter scheme is
particularly important because it both facilitates the individual 
addressability of
the qubits and offers a configuration that could be exploited for performing a large number
of parallel quantum gates (see Fig. 11). 
For both of the configurations the repulsion between electrons is modelled by an
inverse-square interaction $\alpha r^{-2}$ which leads to the same ground-state physics as a
bare Coulomb interaction $r^{-1}$ \cite{Quiroga1}, moreover, such a non-Coulomb
form may actually be more realistic due to the presence of
image charges \cite{Maksym}. 
These configurations are considered in such a way that there is not inter-dot
tunnelling. In such systems we have two combined effects which
are exploited to perform conditional quantum logic gates. First, we have the {\it intra-dot} 
interaction between electrons in
the same QD and their coupling to the nuclear qubits. This interaction produces jumps in 
the relative angular momentum $m$
of the two-electron ground state with increasing $B$. We have shown that these jumps in $m$ 
cause jumps in
the amount of hyperfine splitting in the nuclear spin of the impurity atom, hence providing a
switching mechanism for the nuclear-electron spin transitions \cite{reina2}. Second, we have the
correlation between neighbouring dots, i.e. the
{\it inter-dot} interaction between electrons in different QDs (see Figs. 10(b) and 10(c)). 
As we will see below, this is the
main mechanism responsible for the qubit control given here.

\begin{figure}
%\psdraft
\centerline{\epsfig{file=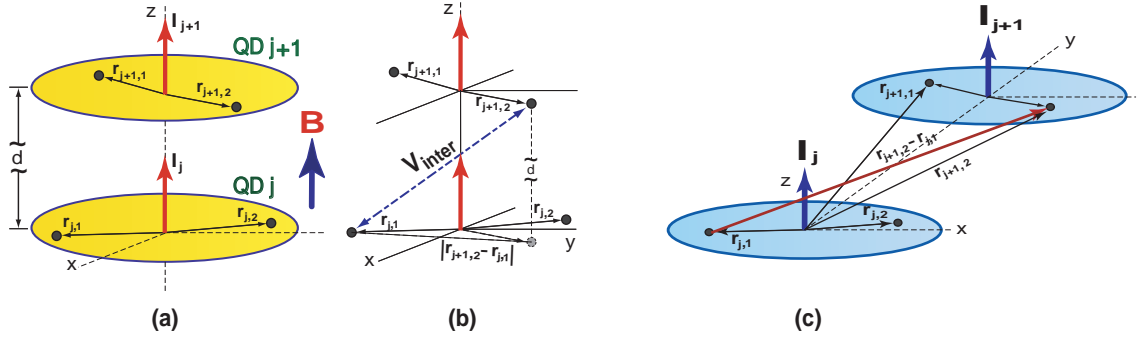,width=15.0cm,height=4.5cm}}
%\vspace{0.3cm}
\caption{\small Schematic of the double QD system. Each QD contains 2 electrons. (a) 
Configuration 1, vertically arranged QDs with the nuclear impurity qubits and the electrons 
in dots $j$ and
$j+1$.  (b) Illustration of 
the {\it inter-dot} correlation
for electrons $(j+1,2)$, and $(j,1)$. Note that the separation between electrons in 
different dots is given in terms of
the distances $|\mbox{\boldmath$r$}_{i+1,\nu}-\mbox{\boldmath
$r$}_{i,\delta}|$ and $d$. The distance $d$ does not correspond with the real scale 
of the system (see text). (c)
Configuration 2, in-plane QDs: schematic of the {\it inter-dot}
correlation for electrons $(j+1,2)$, and $(j,1)$. In this case, all of the dots are 
confined to the $z=0$ plane. The dot
centers, where the nuclear impurities are located, are separated by a constant 
distance $d$. }
\label{figure9}
\end{figure}

\subsection{Hamiltonians and results}

Let us analize the theoretical framework for the switching mechanism and the ability to
tune the electron-nucleus coupling given here. 
The Hamiltonian that models the electron spin-nuclear spin dynamics of the single QDs 
described before,
when  there is no interaction between them, is given 
by $H=H_0+V$ \cite{reina2}, with
\begin{equation}
H_0=H_{2e}+H_{Zeeman}, \ \ V= C\mathop{\displaystyle\sum}
\limits_{\nu=1}^{2}\mbox{\boldmath $I$}\cdot
\mbox{\boldmath$S$}_{\nu}\delta (\mbox{\boldmath $r$}_{\nu})\ \ ,
\end{equation}
where $H_{2e}$ includes the orbital degrees of freedom of the two-electron
QD in a perpendicular magnetic field and $H_{Zeeman}$ corresponds to the
individual electron spins and nuclear spin interaction with the magnetic field.
The Fermi contact
hyperfine coupling of the nuclear spin with the electron spins is expressed by $V$ in
Eq. (16), 
where the
electron-nucleus hyperfine interaction strength is given by
$C={\textstyle{8\pi \over 3}}\gamma_{e}\gamma _{n}\hbar ^{2}|\phi(z=0)|^2$,
with $\phi(z=0)$ the single-electron wavefunction evaluated at the QD plane,
$\gamma _{e}$ ($\gamma _{n}$) is the electronic (nuclear) gyromagnetic ratio
and
$\mbox{\boldmath $S$}_{\nu}$ ($\mbox{\boldmath $I$}$) is the electron
(nuclear) spin.
The electron location in the QD plane is
denoted by the 2D vector
$\mbox{\boldmath $r$}_{\nu}$.
Following
Ref.
\cite{Quiroga1}, $H_{2e}$ splits up into commuting center-of-mass (CM)
motion and relative motion ($rel$) contributions, for which exact
eigenvalues and eigenvectors can be obtained analytically.
The electron-electron interaction only affects the relative motion.
The eigenstates of $H$ can be expressed as linear combinations of
states labeled as $\left|I_{Z};N,M;n,m;S,S_{Z}\right\rangle$, where
$N\;$and $M\;$($n\;$and $m$) are the Landau and angular momentum numbers
for the CM (relative motion) coordinates; $S$ and $S_{Z}$ represent the
total electron spin and its $z$-component, while $I_{Z}$ represents the
$z$-component of the carbon nuclear spin. 
Consider the two-electron system in its ground state, i.e. $N=M=0$, $n=0$;
$m$ determines the orbital symmetry while $S=0,1$ represents the singlet
and triplet spin states respectively.
Neglecting the off-diagonal orbital coupling terms of the hyperfine
interaction $V$,  
the energy associated with the
total Hamiltonian $H$ is
$E=E_{CM}+E_{rel}+E_{spin}$, where $E_{CM}$ ($E_{rel}$) denotes the CM ($rel$)
electron orbital energy contribution and $E_{spin}$ refers to the eigenvalues of the spin
Hamiltonian of the electronic-nuclear system. In the presence of the $B$-field,
the low-lying energy levels all have $n=0$ and $m<0$. The relative angular
momentum $m$ of the two-electron ground state jumps in value with
increasing $B$ (see Refs. \cite{Quiroga1}).
The particular sequence of $m$ values depends on the electron spin because
of the overall antisymmetry of the two-electron wavefunction \cite{Quiroga1}.
For example, only odd values of $m$ arise if the $B$-field is sufficiently
large for the spin wavefunction to be symmetric (the spatial wavefunction
is then antisymmetric). The electron-nucleus coupling depends on the
wavefunction value at the nucleus and hence on $m$. The jumps in $m$ will
therefore cause jumps in the amount of hyperfine splitting in the nuclear
spin of the carbon atom.

The nuclear spin-electron spin effective coupling affecting the
resonance frequency $\omega_{_{NMR}}$ of the  carbon nucleus is given by\begin{equation}
\Delta(m)={\displaystyle{1 \over \pi l^{2}2^{1+\mu_{m}}}} \ \ ,
\end{equation}
where $l=\sqrt{\hbar /m^{\ast }\omega }$ is the effective magnetic length,
the effective frequency is given by $\omega =\sqrt{\omega
_{c}^{2}+4\omega_{0}^{2}}$, $\omega _{c}=eB/m^{\ast }$ is the cyclotron
frequency. The term $\mu_{m}=\left( m^{2}+\frac{\alpha/l_{0}^{2}}{\hbar
\omega _{0}}\right) ^{\frac{1}{2}}$ absorbs the effects of the
electron-electron interaction and $l_{0}=\sqrt{\hbar /m^{\ast }\omega_{0}}$
is the oscillator length. Hence, the effective spin Hamiltonian $H_S$
has the form
\begin{equation}
H_{S}=A(m)\left[(I_{+}S_{-}+I_{-}S_{+})+2I_{Z}S_{Z}\right]
-\gamma _{n}BI_{Z}+\gamma_{e}BS_{Z} \ \ ,
\end{equation}
where $A(m)=\frac{1}{2}C\Delta(m)$ represents a $B$-dependent hyperfine
coupling. We note that the first term of the hyperfine interaction in Eq.
(18)
corresponds to the dynamic part responsible for nuclear-electron flip-flop
spin transitions while the second term describes the static shift of the
electronic and nuclear spin energy levels.

Electrons in the singlet state ($S=0$) are not coupled to the nucleus. In
this case, the nuclear resonance frequency is given by the undoped-QD NMR
signal $\hbar\omega_{_{NMR,0}}=\gamma_n B$. For electron triplet states, the
nuclear
resonance signal corresponds to a transition where the electron spin is
unaffected by a radio-frequency excitation pulse whereas the nuclear spin
experiences a flip. This occurs for the transition between states $\left|
-;1,-1\right\rangle $ and $\left|\Psi\right\rangle=c_1\left|
+;1,-1\right\rangle+c_2\left| -;1,0\right\rangle $. The coefficients $c_1$
and $c_2$ can be obtained analytically by diagonalizing the Hamiltonian
given in Eq. (18). Hence
\begin{equation}
\hbar\omega_{_{NMR}}={\textstyle {3 \over 2}}{\displaystyle{A(m)}+
{\textstyle {1 \over 2}}\left( \gamma _{n}-\gamma
_{e}\right)B}
+{\textstyle {1 \over 2}}\left[\left[
\displaystyle{A(m)}+\left( \gamma _{n}+\gamma _{e}\right) B\right]
^{2}+8A^{2}(m)\right] ^{\frac{1}{2}}.
\end{equation}
Since $\gamma_e>>\gamma_n$, $\hbar\omega_{_{NMR}}\approx\gamma_nB+2A(m)$
which
illustrates the dependence of the NMR signal on the effective $B$-dependent
hyperfine interaction. 

Figure 11(a) shows the effective coupling $\Delta(m)$ between the two-electron
gas and nucleus as a function of the ratio between the cyclotron frequency
and the harmonic oscillator frequency. (The CM is in its ground state).
For silicon, $C/l_0^2=60$ MHz. For $B$-field values where the electron
ground state is a spin singlet ($m$ even) no coupling is present. The
strength of the effective coupling decreases as the $B$-field increases
due to the larger
spatial extension of the relative wavefunction at higher $m$ values, i.e.
the
electron density at the centre of the dot becomes smaller. The
$B$-field provides a very sensitive control parameter for controlling the
electron-nucleus effective interaction. In particular, we note the large
abrupt variation of $\Delta(m)$ for $\frac{\omega_c}{\omega_0}\approx 2.1$
where the electron ground state is performing a transition from a
spin triplet state ($m=1$) to a spin triplet state ($m=3$). This ability to
tune the electron-nucleus coupling underlies the present proposal for an
NMR-based switch.

\begin{figure}
%\psdraft
\centerline{\epsfig{file=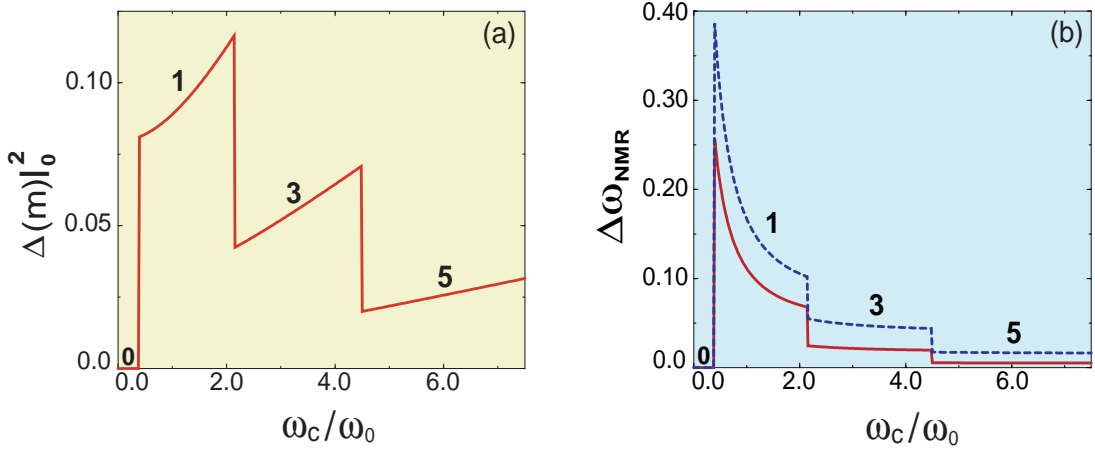,width=14.5cm}}
%\vspace{-0.3cm}
\caption{\small (a) Variation of the electron spin$-$nucleus spin effective coupling
$\Delta(m)$ as a function of $\frac{\omega_c}{\omega_0}$. The center-of-mass motion remains
in its ground state. The electron repulsion strength is given
by $\frac{\alpha /l_{0}^{2}}{\hbar\omega
_{0}}=3.0$. The sequence of transitions is given by $(|m|,S)=\{(0,0), (1,1),
(3,1),(5,1),...\}$. (b) Relative variation of the effective nuclear magnetic resonance 
frequency  of the carbon impurity nucleus. $\frac{\alpha /l_{0}^{2}}{\hbar\omega
_{0}}=3.0$. Solid line corresponds to center-of-mass in the ground
state. Dashed line corresponds to center-of-mass in the first excited state
after absorption of IR light.}
\label{figure11}
\end{figure}

We also give an additional
method for externally controlling the nucleus-electron effective coupling using 
optics \cite{reina2}: in the presence of infra-red (IR) radiation incident on the QD, the CM
wavefunction will be altered since the CM motion absorbs IR radiation. (The
relative motion remains unaffected in accordance with Kohn's theorem). By considering 
the CM transition from the
ground state
$\left| N=0,M=0\right\rangle $ to the excited state $\left|
N=1,M=1\right\rangle$, which becomes the strongest transition in
high $B$-fields, we get the new spin-spin coupling term given by
\begin{equation}
\Delta_{_{CM}}(m)=\left({\displaystyle{1+\mu_{m} \over 2}}
\right)\Delta(m)\ \ .
\end{equation}
Hence the nuclear spin-electron spin coupling is
renormalized by the factor ${\textstyle{1+\mu _{m} \over 2}}$ in the
presence of IR radiation. Figure 11(b) shows the relative variation of $\omega_{_{NMR}}$ with
respect to the undoped QD NMR signal, i.e. $\Delta
\omega_{_{NMR}}=\frac {\omega_{NMR}-
\omega_{NMR,0}}{\omega_{NMR,0}}$ (solid line) as a function of the frequency
ratio $\frac{\omega_c}{\omega_0}$. The jumps in the carbon nucleus
resonance are abrupt, reaching 25\% in the absence of IR radiation.  This
allows a rapid tuning on and off resonance of an incident radio-frequency
pulse. The NMR signal in regions of spin-singlet states remains unaltered.
Moreover, the nuclear spin is being controlled by radio-frequency pulses
which
are externally imposed, thereby offering a significant advantage over
schemes
which need to fabricate and control electrostatic gates near to the qubits,
such as Refs. \cite{kane,privman}. Illuminating the QD with IR
light
will shift the frequencies $\omega_{_{NMR}}$ (see dashed line in Fig. 11(b))
hence
providing further all-optical control of the nuclear qubit. A crucial aspect
of the present proposal is the capability to manipulate individual nuclear
spins. All-optical NMR measurements in semiconductor
nanostructures \cite{nmrgammon,kik} together with local optical probe
experiments are quickly approaching such a level of finesse.

Next, let us consider the situation of a system of $K$ dots which interact with each other: the 
new Hamiltonian associated with this configuration (see Fig. 10) is $H=H_0+V$, with
\begin{equation}
H_0=\mathop{\displaystyle\sum}
\limits_{i=1}^{K}\Big(H_{2e}^{(i)}-\gamma_{n}BI_{i}^{z} + 
\mathop{\displaystyle\sum}\limits_{\nu =1}^{2}\gamma _{e}
BS_{i,\nu}^{z}\Big),\ \ 
V= C\mathop{\displaystyle\sum}
\limits_{i=1}^{K}\mathop{\displaystyle\sum}
\limits_{\nu =1}^{2}\mbox{\boldmath $I$}_{i}\cdot 
\mbox{\boldmath $S$}_{i,\nu }\delta
(\mbox{\boldmath $r$}_{i,\nu})+V_{inter} \ \ ,
\end{equation} 
where $\mbox{\boldmath $S$}_{i,\nu}$
($\mbox{\boldmath $I$}_{i}$) is the spin polarization of electron $\nu$ (nucleus) in dot $i$. The 
location of electron $\nu$ in the $i-$th QD is denoted by
$\mbox{\boldmath $r$}_{i,\nu}$.
The first term in Eq. (21) represents the $i-$th
two-electron QD with a perpendicular $B$-field\footnote{$H_{2e}^{(i)}\equiv H_{CM}^{(i)}+H_{rel}^{(i)}+V_{intra}$ is
given, within  a symmetric gauge, by 
$H_{2e}^{(i)}=\left( \mbox{\boldmath
$P$}_{i,\nu}+2e\mbox{\boldmath $A$}_{i}(\mbox{\boldmath $R$}_{i,\nu})\right) ^{2}/4m^{\ast}+
m^{\ast }\omega _{i,0}^{2} |\mbox{\boldmath $R$}_{i,\nu}|^{2}+
\left(\mbox{\boldmath $p$}_{i,\nu}+\frac{e}{2} \mbox{\boldmath
$A$}_{i}(\mbox{\boldmath
$r$}_{i,\nu})\right) ^{2}/m^{\ast }+
\frac{ m^{\ast }}{4}\omega
_{i,0}^{2}|\mbox{\boldmath $r$}_{i,\nu}|^{2}+
\alpha{|\mbox{\boldmath
$r$}_{i,1}-\mbox{\boldmath
$r$}_{i,2}|^{-2}}$ with $\mbox{\boldmath $R$}_{i,\nu}={\textstyle{1 \over 2}}
\left( \mbox{\boldmath $r$}_{i,1}+\mbox{\boldmath $r$}_{i,2}\right) $, $
\mbox{\boldmath $P$}_{i,\nu}=\mbox{\boldmath $p$}_{i,1}+\mbox{\boldmath $p$}_{i,2}$, $
\mbox{\boldmath $r$}_{i,\nu}=\mbox{\boldmath $r$}_{i,1}-\mbox{\boldmath $r$}_{i,2}$, and 
$\mbox{\boldmath $p$}_{i,\nu}={\textstyle{1 \over 2}}
\left( \mbox{\boldmath $p$}_{i,1}-\mbox{\boldmath $p$}_{i,2}\right).$
}, which includes the {\it intra-dot} interaction 
($V_{intra}$), while the others give the nuclear and the electron-spin Zeeman energies in dot $i$ ($z$ 
indicates the
component of these spin operators). The second term of Eq. (21) give the Fermi contact hyperfine
coupling  of the nuclear
spin of dot $i$ with the electron spin $\nu$ in the same dot and $V_{inter}$ represents the 
{\it inter-dot} interaction between
electrons in neighbouring QDs. The nuclear spin control is performed by the inter-dot 
coupling $V_{inter}$ due to the
interaction between electrons in neighbouring dots. This mechanism (rather than the direct 
dipole-dipole interaction
between the nuclei) is the responsible for the qubit control in the present proposal. In the case 
of configuration 1 (Fig.
10(a)), we have

\begin{equation}
V_{inter}^{(1)}=\mathop{\displaystyle\sum}
\limits_{i=1}^{K}\ \mathop{\displaystyle\sum}
\limits_{\nu,\delta =1}^{2}\frac{\alpha }{|\mbox{\boldmath
$r$}_{i+1,\nu}-\mbox{\boldmath
$r$}_{i,\delta}|^{2}+d^2}\ \ ,
\end{equation} 
where $\mbox{\boldmath
$r$}_{i+1,\nu}-\mbox{\boldmath
$r$}_{i,\delta}\equiv\left([\mbox{$r$}_{i+1,\nu}-\mbox{$r$}_{i,\delta}]_{_{x}},
[\mbox{$r$}_{i+1,\nu}-\mbox{$r$}_{i,\delta}]_{_{y}}\right)$. 
We will assume that the separation between neighbouring QDs is such that
${|\mbox{\boldmath$r$}_{i+1,\nu}-\mbox{\boldmath
$r$}_{i,\delta}|^{2}<d^2}$. This means that the square of the $xy-$plane separation between 
electrons
in neighbouring dots (see Fig. 10(b) for the case of electrons $(j+1,2)$, and $(j,1)$) is 
less than the square of
the vertical separation between such dots ($d^2$), as illustrated in Fig. 10(b). Hence, 
the minimum value
for $d$ is determined by the largest $xy-$projection of electrons in neighbouring dots, 
which roughly
corresponds to the sum of the radii of such dots. The case of configuration 2 (Figs. 10(c) and 11) has

\begin{equation}
V^{(2)}_{inter}=\mathop{\displaystyle\sum}
\limits_{i=1}^{K}\ \mathop{\displaystyle\sum}
\limits_{\nu,\delta =1}^{2}\frac{\alpha }{|\mbox{\boldmath
$r$}_{i+1,\nu}-\mbox{\boldmath
$r$}_{i,\delta}|^{2}}\ \ ,
\end{equation} 
where the in-plane vectors $\mbox{\boldmath $r$}_{i,\epsilon}$ are defined as above. 

\subsubsection{Single-qubit rotations}
Single qubit rotations, e.g. the Hadamard transformation $U_H$, can be performed by rotating the single
nuclear qubit of resonant frequency $\omega_i$ via the application of RF pulses at the appropriate 
frequency for a given
duration and amplitude of the
$B$-field. The coherence time of the nuclear spins is estimated by measuring their $T_1$ and 
$T_2$ relaxation
times, i.e. their nuclear spin-flip relaxation times and the rate of loss of phase coherence 
between the
qubits respectively. In the silicon-based nanostructures considered here, $T_1$
can be  estimated in the 1$-$10 hour range \cite{feher} (for $T<4$ K and $B<1$ T). 
In isotopically purified $^{28}$Si, Si:P linewidths are $<1$ MHz, which gives for $T_2$ 
times greater than 0.5 ms \cite{kane}. In our case, the electrostatically
neutral character of the impurity atom $^{13}$C (and the fact that the
silicon nuclei surrounding it have no nuclear spin) makes the carbon nuclear spin
state very effectively shielded from the environment and hence we would
expect to have far longer $T_2$ times than the (charged) donor nuclei mentioned above.

\subsubsection{The C$-$NOT gate}

The inter-dot interaction potentials given by Eqns. (22) and (23) produce the necessary 
nuclear qubit coupling  for
reliable implementation of the two-qubit gates required for quantum computation. In doing so, 
the Hamiltonian $H$ has
$K=2$ and conditional quantum
dynamics can be performed based on the selective driving of spin resonances of the two 
impurity nuclear qubits, say $I_1$, and
$I_2$, in this system of two coupled QDs (see Fig. 10). The interaction potentials are given by

\begin{equation}
V_{inter}^{(1)}=\mathop{\displaystyle\sum}
\limits_{\nu,\delta =1}^{2}\frac{\alpha }{|\mbox{\boldmath
$r$}_{2,\nu}-\mbox{\boldmath
$r$}_{1,\delta}|^{2}+d^2}\ \ , \ \ 
V^{(2)}_{inter}=\mathop{\displaystyle\sum}
\limits_{\nu,\delta =1}^{2}\frac{\alpha }{|\mbox{\boldmath
$r$}_{2,\nu}-\mbox{\boldmath
$r$}_{1,\delta}|^{2}}\ \ .
\end{equation} 
In these schemes, the orthonormal computation  basis of single
qubits $\left\{\left|0\right\rangle ,\left| 1\right\rangle \right\}$ is represented by the
spin down and up of the impurity nuclei. The QDs do not need to be identical in size. 
The coupling between QDs gives an additional magic number transition as a function of $B$-field 
which can be used for selective switching between
dots, i.e. since the ground state switches back and forth between product states and entangled states
\cite{reina5}, the resonant frequency for transitions between the
states $\left| 0\right\rangle,$ and $\left| 1\right\rangle$ of one nuclear
spin (target qubit)  depends on the state of the other one (control qubit).
In this way, such coupled QDs can be used to generate the conditional C-NOT
gate. The quantum computing scheme proposed here could be easily scalable to 
large quantum information processors:
QDs would be individually addressed via the action of an appropriate $B$-field. This is shown 
in Fig. 12, where the $B_{j,j+1}$ field is assumed to be locally addressing the qubits $j$ and $j+1$ 
from the entire ensemble of dots. Even if the QD array is irregular, one may still be able to perform the
solid-state equivalent of the bulk/ensemble NMR computing recently reported
in Ref. \cite{gershenfeld}. The coupling of the qubits to an external reservoir and the task of controlling
the quantum coherence  of the proposed system are currently being addressed \cite{reina5}. However, and as
we discussed below, due to the exceptionally low decoherence rates of these nuclear qubits, the required RF
pulses  would allow us to perform a sufficient number of single qubit rotations and two-qubit gates 
for realizing ``useful" quantum computing tasks (e.g. the Grover algorithm)  before the system decoheres.

\begin{figure}
%\psdraft
\centerline{\epsfig{file=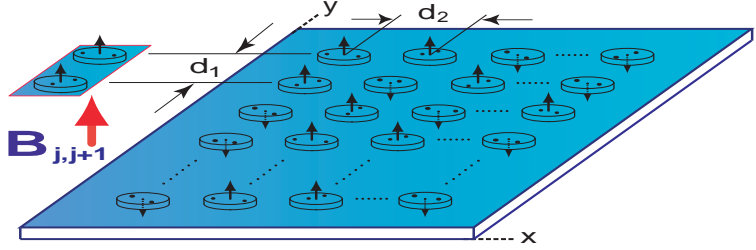,width=10.0cm,height=3.5cm}}
\caption{\small Scaling up Configuration 2 (See Fig. 10 (c)): The $K$ dots of the system 
are confined to the
$z=0$ plane.  Within the entire ensemble of dots, the $B$-field is able 
to locally address:
(a)  single QDs, as in 
the case of one qubit rotations and (b) double dots (e.g. the $j$ and $j+1$ dots in the 
figure) as required 
in the case of two-qubit logic gates.  
The dot centers, where the nuclear impurities are
located,  are separated by constant distances $d_1$ and $d_2$.}
\label{figure12}
\end{figure}

\section{Concluding remarks}

We would like to highlight some aspects of the choice of 
materials and experimental parameters for the implementation of the systems considered here. Regarding the
experimental requirements for building the excitonic setup proposed in Section 2 we point out that hybrid
organic-inorganic nanostructures would be very good  candidates
\cite{Agranovich} since they provide us with large  radius (Wannier-Mott) exciton states in the inorganic
material and small-radius (Frenkel)  exciton states in  the organic one\footnote{There are two models
conventionally used to classify excitons: the small-radius Frenkel exciton model and the large-radius
Wannier-Mott exciton model. Frenkel excitons in organic crystals have radii comparable to the lattice 
constant $a\approx5$A. Wannier excitons in semiconductor quantum wells have large Bohr radii:
$a_B\approx100$A in III-V materials (e.g. GaAlAs) and $a_B\approx30$A in II-VI ones (e.g. ZnSe).}. 
Hence the hybrid material
will be characterized by a radius dominated by their Wannier 
component and by an oscillator strength dominated by their Frenkel
component. This means that the desirable properties of both the organic and the inorganic 
material are brought together to overcome basic limitations which arise if each 
one acts separately. Following recent results \cite{Agranovich}, 
if we consider a system of two or three QDs (as required in the present proposal) of an inorganic 
II-VI material (e.g. the extensively studied ZnSe or ZnCdSe), embedded in bulk-like 
organic crystalline material (e.g. tetracene, perylene, fullerene, PTCDA) where their Frenkel 
and Wannier excitons are in resonance with each other, we would expect a strong hybridization 
between these excitons, which means a greater Wannier exciton delocalization or F\"orster hopping.
To achieve this, the typical distance between QDs should be of the same order 
as their size: In ZnSe, the Bohr radius of the three dimensional Wannier exciton  
$a_B\approx35$A, hence QDs with radii of about 50A will considerably increase the binding energy
of these excitons. If these dots are placed in an organic matrix (as discussed above) separated by a
distance of the same order, we should be able to
observe the entangled states proposed here. There has recently being an experimental observation of photon
antibunching from an artificial atom (a single CdSe/ZnS quantum dot) \cite{Michler}, i.e. the detection of
quantum correlations among photons from a single quantum dot. We note that the statistical properties of
resonance fluorescence from the ensemble of QDs proposed in Section 2 should likewise give raise to a signature
associated with excitonic state entanglement. Theoretical details of this multi-dot
excitonic signature will be reported elsewhere \cite{reina4}.

Regarding the NMR setup of the above section, there may be a natural way to make a quantum dot in silicon
with a single C  atom inside it.   C atoms are known to act as nucleation centers for SiGe quantum dots (see
e.g. Ref.
\cite{schmidt}). Another possibility would be to consider an isolated $^{29}$Si (spin 1/2 and natural
abundance 4.7\%) at the center of a $^{28}$Si based QD.   The isoelectronic character of the impurity is
reinforced but possible purification   procedures could be harder to implement. The more realistic
situation of a non-centered impurity, i.e. when the impurity atom is away from the QD center, will modify
the discontinuity strengths of the electron-nucleus coupling since this coupling is affected by the density
of probability of the CM  wavefunction at the
impurity site \cite{reina2}; however the main effects discussed in the present proposal remain the same.
For a typical $N=2$ electrons QD with 30 $n$m of diameter, lateral confining potential
$\omega_0=8.2\times10^{12}$ s$^{-1}$ ($\hbar\omega_0=5.4$ $m$eV), and low temperatures ($T<$ 1 K)
\cite{ashoori},
we would expect a {\it singlet-triplet} transition $(m,S)=(0,0)\mapsto(1,1)$ at $B\approx1.3$ T, 
or the {\it triplet-triplet}
transition
$(1,1)\mapsto(3,1)$ at
$B\approx6.4$ T. If the harmonic potential is such that $\hbar\omega_0=1.1$ $m$eV 
(see Ref. \cite{ashoori}) the above
transitions would be expected at $B$-fields of 0.3 T and 1.4 T, respectively. For this system, we can
estimate the lower limit of the ``gating time'' $\tau_g$, i.e. the
time for the execution of an individual quantum gate: since the energy splitting
 of the two nuclear qubits, i.e. the value of the energy
difference between the next excited state and the ground singlet and triplet states
of our two-electron system $\Delta E\sim 0.3 $ $m$eV, the lower limit of
$\tau_g$ is
\begin{equation}
\tau_g \gg {\textstyle{ \hbar\over \Delta E}} \sim 1 \ {\rm ps} \,.
\end{equation} 
Therefore, as long as the gating time $\tau_g$ 
is longer than, say, 0.1 $n$s, the QD is 
well isolated, so that the higher excited states can be 
safely neglected, and the gating action can be considered 
adiabatic. The number of
elementary operations that could in principle be performed on a single nuclear
qubit before it decoheres is
${\textstyle{
\tau_{dec}\over
\tau_g}}\approx10^{9}$. This figure of merit should be more than enough to satisfy
the current criteria
for quantum error correction schemes since fault-tolerant 
quantum computation has been shown to be successful if the decoherence time is
$10^4-10^5$ times the gating time. 
Finally, there is the
important issue of the  spin measurements that have to be implemented for either the  
input or the 
readout of single spin
qubits. This process must be rapid enough to avoid decoherence of the  qubits: optical NMR techniques for
reading the input/output of these spin states are currently approaching such a level of finesse
\cite{nmrgammon,kik}.  Other mechanisms for measuring these spin states are also
currently under intensive experimental study \cite{kane1}. 

The solid state NMR proposal given here is not in principle limited to $N=2$ electrons:
generalizations \cite{chak} of the present angular momentum
transitions arise for $N>2$. It was pointed out recently \cite{kouw} that
the  spin configurations in many-electron QDs could be explained in terms of
{\it just} two-electron singlet and triplet states. Therefore, the present
results may occur in QDs with $N>2$. 

It is worth noting that our proposal is not based upon the 
possibility of applying a localized magnetic field to a single quantum dot. 
The procedure to switch the NMR frequency of a single 
nuclear spin is based upon the magic number transitions which can be 
implemented by an extended magnetic field. It is the local hyperfine 
electron-nucleus coupling within each quantum dot which can be 
tuned by such magic number transitions. This is the main point of our 
proposal: the control of the local hyperfine coupling by an extended magnetic 
field may be used to perform single nuclear spin manipulation as well as 
the solid-state equivalent of the 
bulk/ensemble liquid NMR computing (see Ref. \cite{gershenfeld}), for an array  
of either 
identical or non-indentical quantum dots. Similar to 
the NMR liquid experiments, the quantum dot NMR resonance is determined by 
local effects: in our case these are dominated by electron ground-state 
transitions. 
At the present, there is a tremendous 
motivation to perform the 
setup proposed in our work: first, Rabi oscillations have not yet been 
experimentally demonstrated in QDs, and we 
believe that our setup offers an excellent opportunity for doing so. As discussed, the 
ground state energies of QDs with $N$ 
electrons in the presence of magnetic fields, the ones required for our 
proposal, have already been 
experimentallyÊ studied (see e.g. Refs. \cite{ashoori,kouw}). 
Second, there is the possibility of performing quantum logic gates with very 
low decoherence rates. We also would like to note that the magic number transitions
considered here require a relaxation process of the electron
system to achieve the new ground state, i.e. the changes in the
$B$-field (which  change the hyperfine coupling and hence tune the nuclear resonance
frequency) must be done $adiabatically$ to be able to perform the jumps in the angular momentum
quantum number. This electron relaxation
is compatible with the requirement of maintaining the quantum coherence
due to the fact that information is stored in the nuclear spin qubit. 
It is well known that the nuclear spin relaxation times are several orders of magnitude
longer than electron relaxation times. Therefore, electrons in the quantum
dot can evolve to a new ground state before any environmentally-induced
contamination affects the nuclear spin state.  

To summarize, we have shown that semiconductor nanostructures can be exploited in order to realize 
{\it all-optical} quantum entanglement schemes, even in the presence of noisy environments. A scheme for quantum
teleportation of excitonic states has also been proposed. In addition, we have presented a solid state
NMR-based mechanism for performing reliable quantum computation.
\\

\noindent {\bf Acknowledgements.} The authors acknowledge the suppport of the Colombian 
government agency for 
science and technology, COLCIENCIAS.

\bibliographystyle{plain}

\end{document}